\documentclass[prb,preprintnumbers,amsmath,amssymb,superscriptaddress]{revtex4}

\usepackage{graphicx}
\usepackage{bm}

\begin{document}

\title{Computation of dynamical correlation functions of Heisenberg chains: \\
the gapless anisotropic regime}

\author{Jean-S\'ebastien Caux} 
\author{Rob Hagemans}
\affiliation{Institute for Theoretical Physics, University of
  Amsterdam, 
  1018 XE Amsterdam, The Netherlands.}
\author{Jean-Michel Maillet} 
\affiliation{Laboratoire de Physique, \'Ecole Normale Sup\'erieure de Lyon, 
  69364 Lyon CEDEX 07, France.}

\date{\today}

\begin{abstract}
We compute all dynamical spin-spin correlation functions for the spin-$1/2$ $XXZ$ anisotropic Heisenberg model
in the gapless antiferromagnetic regime, using numerical sums of exact determinant representations for form factors of
spin operators on the lattice.  
Contributions from intermediate states containing many particles and string (bound) states are included.  We
present modified determinant representations for the form factors valid in the general case with string solutions to the
Bethe equations.
Our results are such that the available sum rules are saturated to high precision.
We Fourier transform our results back to real space, allowing us in particular to make a comparison with known exact formulas
for equal-time correlation functions for small separations in zero field, and with
predictions for the zero-field asymptotics from conformal field theory.
\end{abstract}

\maketitle

\section{Introduction}
In recent years, enormous progress has been made in the quest to overcome 
an important and long-standing limitation of the Bethe Ansatz \cite{BetheZP71} framework in the theory of
integrable models: the inability to compute correlation functions. 
Most efforts have been focused on  
the Heisenberg spin chain \cite{JimboBOOK,KitanineNPB554,KitanineNPB567,Boosh0412191},
although similiar sets of results could in principle be obtained for other integrable models 
formulated via the Algebraic Bethe Ansatz \cite{KorepinBOOK}.
In the particular case of the $XXZ$ model, matrix elements of any
local operator between two Bethe states can now be written as 
determinants of matrices, whose elements are known analytic functions of the rapidities of the eigenstates involved \cite{KitanineNPB554}.  
Used in conjunction with formulas for
eigenstate norms \cite{KorepinCMP86}, this yields exact expressions for form factors on the lattice,
thereby permitting in principle the computation of dynamical correlation functions.

In this paper, we concentrate on the anisotropic Heisenberg spin-1/2 chain in a magnetic field, with
Hamiltonian
\begin{eqnarray}
H \!=\! J \!\sum_{j = 1}^N \!\left[ S^x_j S^x_{j+1} \!+\! S^y_j S^y_{j+1} \!+\! \Delta \!\left(S^z_j S^z_{j+1}\!-\!\frac{1}{4}\right) \!-\! h S^z_j \right]
\label{XXZ}
\end{eqnarray}
and periodic boundary conditions.  The anisotropy parameter $\Delta$ can take on any real value without destroying
the applicability of the Bethe Ansatz or of the present method.  We will however consider
here only the quantum critical regime $-1 < \Delta \leq 1$.  

Our interest lies in space- and time-dependent spin-spin correlation functions.  Continuing recent work by two of us \cite{Caux0502365}, 
we numerically obtain the dynamical spin-spin structure factor, defined as the Fourier transform of the connected spin-spin correlation
function
\begin{eqnarray}
S^{a\bar{a}} (q, \omega) = \!\frac{1}{N} \!\sum_{j, j'=1}^N \!e^{i q(j - j')} \!\int_{-\infty}^{\infty} dt e^{i\omega t} 
\langle S^a_j(t) S^{\bar{a}}_{j'}(0) \rangle_c
\label{Structurefactor}
\end{eqnarray}
where $a = z, \pm$.  This quantity is of fundamental importance in many settings, and is used for example in the 
quantitative description of inelastic neutron scattering data for quasi-one-dimensional systems
\cite{KenzelmannPRB65,StonePRL91,ZaliznyakPRL93,LakeNature4} to more theoretical questions relating to 
quantum entanglement \cite{JinJPA69}.  In any case, the main interest is based on the fact that the Heisenberg
spin chain is a strongly coupled quantum system with fractionalized (spinon-like and higher) excitations, for
which it is possible to obtain nonperturbative information.

The use of Bethe Ansatz-based methods to obtain quantitative results on correlation functions, in view
of the long history of integrability, is rather recent.  In the case of zero field, much progress has been
possible for the model in the gapped phase \cite{JimboBOOK}, and for the isotropic limit, for which both two-spinon
\cite{BougourziPRB54,KarbachPRB55} and four-spinon \cite{AbadaNPB497} contributions were obtained analytically.  The nonzero
field problem remains intractable with this approach.  However, summations over intermediate states can be
performed numerically once the analytical expressions for the form factors are known.  Using this strategy,     
the spin-spin correlations (both longitudinal and transverse) for the isotropic Heisenberg model in a field were studied
at fixed momentum $q$ in [\onlinecite{BiegelEPL59}], and the transverse ones for $XXZ$ at zero field and at $q = \pi$ in [\onlinecite{BiegelJPA36}].
Two-particle contributions to the longitudinal structure factor for the $XXZ$ chain in a field at
$q = \pi/2$ were also studied in [\onlinecite{SatoJPSJ73}].
Multiparticle contributions and extension to the full Brillouin zone was done in [\onlinecite{Caux0502365}].
In these, it is observed that summing over a relatively small subset of intermediate states is sufficient to obtain good
precision, as demonstrated by using sum rules.  The present paper has three main further objectives:
to generalize the form factor determinant formulae
to intermediate states containing complex rapidities (in the form of strings) present in the Bethe Ansatz for the $XXZ$ chain, 
to study the general field dependence of the correlation functions, 
and to obtain the real space- and time-dependent
correlators.
 
The plan of the paper is as follows.  We begin in section \ref{Setup} by recalling a few basic facts concerning the Bethe Ansatz
for the $XXZ$ model, and the lattice form factors for single spin operators.  In section \ref{Strings}, we discuss string solutions to the
Bethe equations for excited states, and obtain modified determinant representations for the form factors
of local spin operators involving such states.  Section \ref{Plots} contains the results of our numerical evaluations
of the form factor sums for dynamical spin-spin correlation functions for two example choices of anisotropy.
In section \ref{Fourier}, we use our results to compute the space- and time-dependent correlators.  This allows us
in particular to obtain equal-time correlators for any lattice separation.  Comparison is made with known exact
results in zero field for small lattice separation.
We also compare our results to predictions from conformal field theory for the asymptotics in zero field.
Our conclusions and outlook are
collected in section \ref{Conclusion}.

\section{Setup}
\label{Setup}
In this section, we briefly review all elements necessary for the computation of the spin-spin correlation functions 
of the anisotropic Heisenberg model.
The exact solution through the Bethe Ansatz for model (\ref{XXZ}) is well-known (see [\onlinecite{KorepinBOOK,TakahashiBOOK}]
and references therein).  
The reference state is taken to be the state with all spins up, 
$|0\rangle = \otimes_{i = 1}^N |\uparrow\rangle_i$.  Since the total magnetization commutes with the Hamiltonian, 
the Hilbert space separates into subspaces of fixed magnetization,
determined from the number of reversed spins $M$.  We take the number of sites $N$ to be even, and $2M \leq N$,
the other sector being accessible through a change in the reference state.

Eigenstates in each subspace are completely characterized for $2M \leq N$ by a set of rapidities $\{\lambda_j\}$, 
$j = 1, ...,M$, solution to the Bethe equations
\begin{eqnarray}
\left[ \frac{\sinh (\lambda_j + i\zeta/2)}{\sinh(\lambda_j - i\zeta/2)} \right]^N = \prod_{k \neq j}^M 
\frac{\sinh(\lambda_j - \lambda_k + i\zeta)}{\sinh(\lambda_j - \lambda_k - i\zeta)}, \hspace{1cm} j = 1, ..., M
\label{BAE}
\end{eqnarray}
where $\Delta = \cos \zeta$.  In view of the periodicity of the $\sinh$ function in the complex plane, we can
restrict the possible values that the rapidities can take to the strip $-\pi/2 < \mbox{Im} \lambda \leq \pi/2$, or
alternately define an extended zone scheme in which $\lambda$ and $\lambda + i\pi {\mathbb Z}$ are identified.

A more practical version of the Bethe equations is obtained by writing them in logarithmic form,
\begin{eqnarray}
\mbox{atan} \!\!\left[ \frac{\tanh(\lambda_j)}{\tan(\zeta/2)} \right] \!\!-\! \frac{1}{N} \!\sum_{k = 1}^M 
\mbox{atan} \!\!\left[\frac{\tanh(\lambda_j - \lambda_k)}{\tan \zeta}\right] \!\!=\! \pi \frac{I_j}{N}.
\label{BAE_log}
\end{eqnarray}
Here, $I_j$ are distinct half-integers which can be viewed as quantum numbers:  
each choice of a set $\{I_j\}$, $j = 1, ...,M$ (with $I_j$ defined mod$(N)$)
uniquely specifies a set of rapidities, and therefore an eigenstate.  The energy of a state is given
as a function of the rapidities by
\begin{eqnarray}
E = J \sum_{j = 1}^M \frac{-\sin^2 \zeta}{\cosh 2\lambda_j - \cos \zeta} - h(\frac{N}{2} - M), 
\end{eqnarray}
whereas the momentum has a simple representation in terms of the quantum numbers,
\begin{eqnarray}
q = \sum_{j = 1}^M i \ln \left[\frac{\sinh(\lambda_j + i\zeta/2)}{\sinh(\lambda_j - i\zeta/2)}\right]
= \pi M + \frac{2\pi}{N}\sum_{j = 1}^M I_j \hspace{0.5cm} \mbox{mod} \hspace{0.2cm}2\pi.
\label{Ep}
\end{eqnarray}
The ground state is given by $I_j^0 = -\frac{M+1}{2} + j$, $j = 1, ...,M$, and all excited states are in principle obtained from 
the different choices of sets $\{ I_j \}$.  

To study dynamics, some ingredients have to be added to the Bethe Ansatz:  the matrix elements of spin
operators between eigenstates (form factors).  
In terms of form factors for the Fourier-transformed spin operators $S^a_q = \frac{1}{\sqrt{N}} \sum_{j=1}^N e^{i q j} S^a_j$, 
the structure factor (\ref{Structurefactor}) can be written as a sum
\begin{eqnarray}
S^{a\bar{a}} (q, \omega) = 2\pi \sum_{\alpha} |\langle GS | S_q^a | \alpha \rangle |^2 \delta(\omega -\omega_{\alpha})
\label{S_zz_ff}
\end{eqnarray}
over the whole set of contributing intermediate eigenstates $|\alpha \rangle$.  For the longitudinal structure factor,
this is the set of all states with the same number of overturned spins as the ground state in the chosen magnetization
subsector, excluding the ground
state itself (connected correlator).  For the transverse structure factor $S^{-+}$, it is the set of all states with
one less overturned spin.  $\omega_{\alpha}$ is the energy difference of state $|\alpha \rangle$ with the ground state.
Each term in (\ref{S_zz_ff}) can be obtained \cite{KitanineNPB554} as a product of determinants of 
specific matrices, which are fully determined for given bra and ket eigenstates by a knowledge of the corresponding sets of rapidities.

The form factor between two eigenstates (with $M$ reversed spins, and rapidities $\{\mu \}$, $\{\lambda \}$) for the $S^z$ operator is given by
\cite{KitanineNPB554}
\begin{eqnarray}
|\langle \{\mu\}|S_q^z|\{\lambda\}\rangle|^2 = 
\frac{N}{4} \delta_{q, q_{\{\lambda\}} - q_{\{\mu\}}}\prod_{j=1}^M \left|\frac{\sinh(\mu_j - i\zeta/2)}{\sinh(\lambda_j - i\zeta/2)}\right|^2
\prod_{j > k = 1}^M \left|\sinh^2(\mu_j - \mu_k) + \sin^2 \zeta\right|^{-1} \times \nonumber \\
\times \prod_{j > k = 1}^M \left|\sinh^2(\lambda_j - \lambda_k) + \sin^2 \zeta\right|^{-1}
\frac{\left|\det[{\bf H}(\{\mu\}, \{\lambda\}) - 2{\bf P}(\{\mu\}, \{\lambda\})]\right|^2}
{\left|\det {\bf \Phi} (\{\mu\}) \det {\bf \Phi} (\{\lambda\}) \right|}
\label{Szformfactor}
\end{eqnarray}
where the matrices $H$ and $P$ are defined as
\begin{eqnarray}
{\bf H}_{ab} (\{\mu\}, \{\lambda\}) = \frac{1}{\sinh(\mu_a - \lambda_b)} \left[ \prod_{j\neq a}^M \sinh(\mu_j - \lambda_b - i\zeta)
- \left[\frac{\sinh(\lambda_b + i\zeta/2)}{\sinh(\lambda_b - i\zeta/2)}\right]^N \prod_{j \neq a}^M \sinh(\mu_j - \lambda_b + i\zeta)\right],
\end{eqnarray}
\begin{eqnarray}
{\bf P}_{ab} (\{\mu\}, \{\lambda\}) = \frac{1}{\sinh^2\mu_a + \sin^2 \zeta/2} \prod_{k = 1}^M \sinh(\lambda_k - \lambda_b - i\zeta), 
\end{eqnarray}

Similarly, the form factor for the $S^-$ operator between eigenstates with $M$ and $M-1$ reversed spins (respectively
having sets of rapidities given by $\{\mu \}$, $\{\lambda \}$) is \cite{KitanineNPB554}
\begin{eqnarray}
|\langle \{\mu\}|S_q^-|\{\lambda\}\rangle|^2 = 
N \delta_{q, q_{\{\lambda\}} - q_{\{\mu\}}}
|\sin \zeta| \frac{\prod_{j=1}^{M} \left|\sinh(\mu_j - i\zeta/2)\right|^2}{\prod_{j = 1}^{M-1} \left|\sinh(\lambda_j - i\zeta/2)\right|^2}
\prod_{j > k = 1}^{M} \left|\sinh^2(\mu_j - \mu_k) + \sin^2 \zeta\right|^{-1} \times \nonumber \\
\times \prod_{j > k = 1}^{M-1} \left|\sinh^2(\lambda_j - \lambda_k) + \sin^2 \zeta\right|^{-1}
\frac{\left|\det{\bf H^-}(\{\mu\}, \{\lambda\}) \right|^2}
{\left|\det {\bf \Phi} (\{\mu\}) \det {\bf \Phi} (\{\lambda\}) \right|}
\label{Smformfactor}
\end{eqnarray}
in which the $H^-$ matrix is defined as
\begin{eqnarray}
{\bf H}_{ab}^- (\{\mu\}, \{\lambda\}) = \frac{1}{\sinh(\mu_a - \lambda_b)} \left[ \prod_{j\neq a}^{M} \sinh(\mu_j - \lambda_b - i\zeta)
- \left[\frac{\sinh(\lambda_b + i\zeta/2)}{\sinh(\lambda_b - i\zeta/2)}\right]^N \prod_{j \neq a}^{M} \sinh(\mu_j - \lambda_b + i\zeta)\right], 
b < M, \nonumber \\
{\bf H}_{a M}^- (\{\mu\}, \{\lambda\}) = \frac{1}{\sinh^2\mu_a + \sin^2 \zeta/2}. \hspace{6cm}
\label{Hm}
\end{eqnarray}

The norm of the eigenstates appears in the above expressions in the form of the determinant of the Gaudin matrix
\cite{GaudinBOOK,KorepinCMP86}
\begin{eqnarray}
{\bf \Phi}_{ab} (\{\lambda\}) = \delta_{ab} \left[ N\frac{\sin\zeta}{\sinh^2 \lambda_a + \sin^2 \zeta/2} - \sum_{k \neq a} 
\frac{\sin 2\zeta}{\sinh^2 (\lambda_a - \lambda_k) + \sin^2 \zeta} \right] 
+ (1 - \delta_{ab}) \frac{\sin 2\zeta}{\sinh^2 (\lambda_a - \lambda_b) + \sin^2 \zeta}.
\end{eqnarray}

\section{Strings}
\label{Strings}
Although the ground state of the $XXZ$ chain is characterized by real rapidities, this is not true for all
excited states.  It has been known ever since the original work of Bethe \cite{BetheZP71} that there exist
sets of complex rapidities satisfying the Bethe equations.  Typically, complex rapidities come in
conjugate pairs or more elaborate structures involving higher numbers of elements.  
These complex rapidity groupings represent bound states of overturned spins, with spatial extent inversely related to
the imaginary parts.  At low densities (so at high magnetic fields), most solutions take the form of
so-called strings, in which a number of rapidities share a real center but have regularly-spaced imaginary parts.   
For the $XXZ$ model, a classification of the possible types of string solutions 
has been conjectured in [\onlinecite{TakahashiPTP46}], and further justified in [\onlinecite{HidaPLA84,FowlerPRB24}].
This classification depends sensitively on the value of the anisotropy parameter $\Delta$, and string state counting
is traditionally used to argue for the completeness of the string state basis both for the isotropic magnet
\cite{TakahashiPTP46_1,GaudinBOOK} as for the anisotropic one \cite{KirillovJPA30}.  

Our principal aim in this section is to adapt the determinant formulae for form factors and state norms 
to the case where one of the eigenstates involved contains string solutions to the Bethe equations.  This is
a necessary procedure, as both the Bethe equations and the determinant formulae for form factors suffer from
degenerate limits in the presence of strings, which have to be analytically 
dealt with by hand.  Let us however
begin by reviewing the string structure, and setting out our conventions and notations.

Following Takahashi and Suzuki \cite{TakahashiPTP46}, we take $\zeta/\pi$ to be a real number between $0$ and $1$, which is expressed as a 
continued fraction of real positive integers as
\begin{eqnarray}
\frac{\zeta}{\pi} = \frac{1|}{|\nu_1} + \frac{1|}{|\nu_2} + ... \frac{1|}{|\nu_l}, \hspace{1cm} 
\nu_1, ..., \nu_{l-1} \geq 1, \nu_l \geq 2.
\end{eqnarray} 
For large $N$, rapidities congregate to form strings centered either on the real line (for positive parity strings) 
or on the axis $i\pi/2$ (for negative parity strings),
\begin{eqnarray}
\lambda^{n_j, a}_{\alpha} = \lambda^{n_j}_{\alpha} + i\frac{\zeta}{2} (n_j + 1 - 2a) + i\frac{\pi}{4} (1 - v_j) + i \delta^{n_j, a}_{\alpha}, 
\hspace{1cm} a = 1, ..., n_j
\label{string_rapidities}
\end{eqnarray}
where the allowable lengths $n_j$ and parities $v_j = \pm 1$ are to be determined.  In a string
configuration, the parameters $\delta^{n_j, a}_{\alpha}$ are exponentially suppressed with system size.

The classification of allowable string types in the thermodynamic limit proceeds according to the following
algorithm.  First, the positive integer series
$y_{-1}, y_0, y_1, ..., y_l$ and $m_0, m_1, ..., m_l$ are defined as
\begin{eqnarray}
y_{-1} = 0, ~~y_0 = 1, ~~y_1 = \nu_1, ~~\mbox{and}~~ y_i = y_{i-2} + \nu_i y_{i-1}, ~~i = 2, ..., l, \hspace{1cm}
m_0 = 0, ~~m_i = \sum_{k = 1}^i \nu_k.
\end{eqnarray}
Lengths and parities are then given by (our conventions here have
the advantage of giving a proper ordering of string lengths, $n_j > n_k$, $j > k$)
\begin{eqnarray}
n_j = y_{i-1} + (j - m_i)y_i, ~~v_j = (-1)^{\lfloor (n_j - 1)\zeta/\pi \rfloor}, \hspace{1cm} m_i \leq j < m_{i+1}.
\end{eqnarray}
The total number of possible strings is $N_s  = m_l + 1$, and the index $j$ runs over the set $1, ..., N_s$.  
The real parameters $\lambda_{\alpha}^{n_j}$ represent the centers
of strings with length $n_j$ and parity $v_j$, and are hereafter noted as $\lambda^j_{\alpha}$, $\alpha = 1, ..., M_j$, 
where $M_j$ is the number of strings of length $n_j$ in the eigenstate under consideration.  We therefore have
the constraint $\sum_{j = 1}^{N_s} n_j M_j = M$.

In a string configuration, many factors appearing in the Bethe equations become of the indeterminate form $\delta/\delta$.  
Remultiplying
(\ref{BAE}) for each member of a particular string gets rid of these factors, and allows one to rewrite the whole
set of Bethe equations in terms of a reduced set involving only the string centers $\lambda_{\alpha}^j$.
Doing this, one finds (for $N$ even) the reduced set of Bethe-Takahashi equations \cite{TakahashiPTP46}
\begin{eqnarray}
\bar{e}_j^N (\lambda_{\alpha}^j) = (-1)^{M_j-1}\prod_{(k,\beta) \neq (j,\alpha)} \bar{E}_{jk} (\lambda_{\alpha}^j - \lambda_{\beta}^k),
\label{BAE_strings}
\end{eqnarray}
where 
\begin{eqnarray}
\bar{e}_j(\lambda) = -v_j\frac{\sinh(\lambda + i\frac{\pi}{4}(1 - v_j) + i n_j\zeta/2 )}{\sinh(\lambda + i\frac{\pi}{4}(1 - v_j) - i n_j\zeta/2)}, 
\hspace{1cm}
\bar{E}_{jk}(\lambda) = \bar{e}_{|n_j-n_k|}^{1-\delta_{n_j n_k}} (\lambda) \bar{e}_{|n_j-n_k|+2}^2(\lambda) ... 
\bar{e}_{n_j+n_k-2}^2 (\lambda) \bar{e}_{n_j+n_k}(\lambda).
\end{eqnarray}
For the state classification and computation, it is preferable to work with the logarithmic form
\begin{eqnarray}
N\theta_j(\lambda^j_{\alpha}) - \sum_{k=1}^{N_s} \sum_{\beta=1}^{M_k} \Theta_{jk}(\lambda^j_{\alpha} - \lambda^k_{\beta}) = 2\pi I^j_{\alpha}
\label{BAE_log_strings}
\end{eqnarray}
where $I^j_{\alpha}$ is integer if $M_j$ is odd, and half-integer if $M_j$ is even.
The dispersion kernels and scattering phases appearing here are 
\begin{eqnarray}
\theta_j (\lambda) &=& 2 v_j ~\mbox{atan} \left[(\tan n_j\zeta/2)^{-v_j}\tanh \lambda \right], \nonumber \\
\Theta_{jk}(\lambda) &=& (1-\delta_{n_j n_k}) \theta_{|n_j-n_k|}(\lambda) + 2\theta_{|n_j-n_k|+2}(\lambda) + ... + 
2\theta_{n_j+n_k-2} (\lambda) + \theta_{n_j+n_k} (\lambda).
\end{eqnarray}
The energy and momentum of a string are given by
\begin{eqnarray}
E^{j}_{\alpha} = - J \frac{\sin \zeta \sin n_j \zeta}{v_j \cosh 2\lambda^j_{\alpha} - \cos n\zeta}, 
\hspace{1cm} p_0 (\lambda^j_\alpha) = i \ln \left[\frac{\sinh(\lambda^j_{\alpha} 
+ i \frac{\pi}{4}(1 - v_j) + in_j \zeta/2)}
{\sinh(\lambda^j_{\alpha} + i \frac{\pi}{4}(1 - v_j) - in_j \zeta/2)} \right]
\end{eqnarray}
so the total momentum is again expressible in terms of the string quantum numbers as
\begin{eqnarray}
q = \pi T_{v = +} + \frac{2\pi}{N} \sum_{j=1}^{N_s}\sum_{\alpha = 1}^{M_j} I_{\alpha}^j 
\hspace{0.5cm}\mbox{mod} 2\pi
\end{eqnarray} 
in which $T_{v=+}$ is the total number of excitations with positive parity.

It is known that the string hypothesis is not valid in absolute generality, meaning that not all eigenstates are
described by a Bethe wavefunction with rapidities arranged in string patterns.  For example, in the sector with
two down spins of the isotropic antiferromagnet, it was demonstrated \cite{EsslerJPA25} that, for large lattice
size $N$, there exist $\mbox{O} (\sqrt{N})$ pairs of extra real solutions replacing the corresponding expected
two-strings (the total number of two-strings scaling like $N$).  A fraction $1/\sqrt{N}$ of two-string states in
the two down spin sector thus fail to be captured by the string hypothesis.  For the anisotropic $XXZ$ model,
again in the two down spin sector, it was argued \cite{IlakovacPRB60} that the number of violations of the
string hypothesis remains finite for any $\Delta < 1$ as $N$ goes to infinity.  Non-string states for the
$XXZ$ model were also studied in [\onlinecite{FujitaJPA36}].  In both cases, however, the
total number of solutions to the Bethe equations given on the basis of the string hypothesis correctly reflects
the dimensionality of the Hilbert space.  If a string goes missing, it is replaced by a pair of real rapidities.

The problem of counting the number of violations of the string hypothesis for larger numbers of overturned
spins remains generally open.  The one limit in which analytic progress can be made is at
zero field, where $M = N/2$.  There, it is possible to show \cite{WoynarovichJPA15,BabelonNPB220} that, in
the thermodynamic limit, the lowest-lying excited states involving complex rapidities can only take the form
of either complex conjugate pairs or self-conjugate quartets.  Strings of length higher than two therefore
seem to be excluded.  We therefore expect to see a breakdown of higher-than-two strings as the magnetization
decreases from its saturation value ({\it i.e.} the limit in which strings exist) down to zero.

Since we do only a partial trace over the set of all intermediate states, 
including only a minuscule proportion of the whole set, we do not concern ourselves here 
with non-string states.  This will have as a consequence that our results will be ``at their worst'' at zero field,
where we only sum up states including two-string excitations, 
but will be essentially exact at higher fields.  For each string state, we check explicitly that the deviations
$\delta$ in (\ref{string_rapidities}) are exponentially small, and therefore only include these contributions if the
string hypothesis is explicitly verified.
In any case, we observe in general that strings of length higher than two give
negligible contributions to the correlation functions.  
We therefore use the string hypothesis here as a general basis for our state 
classification scheme.  The quality of our results will be confirmed later on with the use of sum rules.

Our terminology for the state classification is as follows.  
First, we define a base as the set $\{ M_k \}$, $k = 1, ..., N_s$ of numbers of each string
type present in an eigenstate.  Many eigenstates share a same base, and these are obtained by choosing different quantum
number configurations $\{ I_{\alpha}^j \}$, $j = 1, ..., N_s$, $\alpha = 1, ..., M_j$.  The base itself defines
the requirements on these quantum numbers.  First, $I_{\alpha}^j$ must be integer if $M_j$ is odd, and half-integer if $M_j$ is even.
Second, they should be contained in the set $\{ -I_{\infty}^j, -I_{\infty}^j + 1, ..., I_{\infty}^j \}$, where
\begin{eqnarray}
I^j_{\infty} = \lfloor \frac{1}{2\pi} | N \theta_j (\infty) - \sum_{k = 1}^{N_s} (M_k - \delta_{jk}) \Theta_{jk} (\infty) | \rfloor_j
\end{eqnarray}
in which the notation $\lfloor a \rfloor_j$ means that we take the highest integer or half-integer less than or equal to the argument $a$, 
depending on the parity of $M_j$.  The base choice completely specifies the right-hand side of this equation, and therefore the
limits $I_{\infty}^j$.  

The total number of configurations sharing a given base is then given by 
$D_{\{ M \}} = \prod_j \left( \begin{array}{c} 2I_{\infty}^j + 1 \\ M_j
\end{array} \right) $.  The total number of possible string states is then upon first examination given by
summing this over all possible bases, $D = \sum_{\{ M \}} D_{\{ M \}}$.  This line of reasoning is used to argue for
the completeness of the string states.
However, we find that within a given base, there exist a number of inadmissible configurations.  
These correspond
to configurations that are parity-symmetric, $\{ -I_{\alpha}^j \} = \{ I_{\alpha}^j \}$ and
that have a vanishing quantum number for at least one higher string,   
$I_{\alpha}^j = 0$ with even $n_j$, or that have two vanishing quantum numbers $I_{\alpha}^j, I_{\beta}^k$, $j \neq k$, 
with $n_j, n_k$ odd.  In the first case, there are rapidities exponentially close to $i\zeta/2$
(which represents an ill-defined limit when obtaining the Bethe equations), and in the second at least two exponentially close 
rapidities, meaning a wavefunction improperly defined by the straightforward Bethe Ansatz.  The
correct eigenstates should thus presumably be obtained by a proper limiting procedure.
In view of the correctness of the string hypothesis 
state counting \cite{KirillovJPA30}, these must correspond to non-string states.  We will readdress the question
of state counting and non-string states in a separate publication.  For our purposes here, however, completeness is
an academic question, since we only need to do partial traces over intermediate states (albeit including strings).  

The presence of strings in the distribution of the rapidities $\{ \lambda \}$ of an eigenstate renders the 
determinants used in (\ref{Szformfactor}), (\ref{Smformfactor}) degenerate, by which we mean that columns of the
$H, P, H^-$ and $\Phi$ matrices become equal to leading order in the string deviations $\delta$.
The matrix determinants must therefore be expanded to higher order, and column/row manipulations used to extract
the first nonvanishing contribution.  Overall, the form factors have proper limits.  

It is straightforward to show that the norm of the eigenstate with strings is given by the determinant of a reduced Gaudin matrix, 
\begin{eqnarray}
|\det \Phi (\{ \lambda \})| = \left[ \prod_{j | n_j > 1}\prod_{\alpha}\prod_{a=1}^{n_j - 1} |(\delta_{\alpha}^{n_j, a} - \delta_{\alpha}^{n_j, a+1})|\right]
\left[ |\det \Phi^{(r)}| + \mbox{O} (\delta) \right]
\label{reduced_matrix}
\end{eqnarray} 
of dimension equal to the number of strings in the state $\{ \lambda \}$.  Its matrix elements are defined as
\begin{eqnarray}
\Phi^{(r)}_{(j, \alpha) (k,\beta)} = \delta_{jk}\delta_{\alpha \beta} \left( N \frac{d}{d\lambda^j_{\alpha}} \theta_j(\lambda^j_{\alpha})
- \sum_{(l,\gamma) \neq (j, \alpha)} \frac{d}{d\lambda^j_{\alpha}} \Theta_{jl} (\lambda^j_{\alpha} - \lambda^l_{\gamma}) \right)
+ (1-\delta_{jk} \delta_{\alpha \beta}) \frac{d}{d\lambda^j_{\alpha}} \Theta_{jk} (\lambda^j_{\alpha} - \lambda^k_{\beta}).
\end{eqnarray}

In the case where the eigenstate described by the set $\{\lambda \}$ contains strings and the one described by $\{ \mu \}$ 
does not, we can write explicit regular formulas for the elements of the matrices appearing in the longitudinal and transverse
form factors.  We find that the $H$ and $P$ matrix determinants for the longitudinal 
form factor can be reduced, in the presence of strings, to the determinants of reduced matrices up to corrections of order of the
string deviations $\delta$,
\begin{eqnarray}
|\det [H - 2P] | = | \det [H^{(r)} - 2P^{(r)}]| + \mbox{O}(\delta).
\end{eqnarray}
The matrix components of $H^{(r)}$ and $P^{(r)}$, 
for matrix elements where $\lambda_b$ is part of an $n$-string, are given by
\begin{eqnarray}
H_{ab}^{(r)} &=& K^{(n)}_{ab}, \hspace{1cm} P_{ab}^{(r)} = 0, 
\hspace{1cm} b = 1, ..., n-1, \nonumber \\
H_{ab}^{(r)} &=& N^{(n)} \sum_{j=0}^n \frac{G^{(n)}_j G^{(n)}_{j+1}}{F^{(n)}_j F^{(n)}_{j+1}}
\left( [\delta_{j,0} + \delta_{j,n} - 1] L^{(n)}_{aj} + [\delta_{j,0} + \delta_{j,n}] K^{(n)}_{aj}\right), 
\nonumber \\
P_{ab}^{(r)} &=& N^{(n)} \frac{1}{\sinh^2 \mu_a + \sin^2 \zeta/2} \sum_{j=1}^n \frac{G^{(n)}_j}{F^{(n)}_j}, \hspace{1cm} b = n,
\end{eqnarray}
where 
\begin{eqnarray}
F^{(n)}_b &=& \prod_{k \neq b} \sinh\left[ \lambda_k - \lambda_b\right], \hspace{1cm}
G^{(n)}_b = \prod_{j=1}^M \sinh \left[ \mu_j - \lambda_b\right], \nonumber \\
K^{(n)}_{ab} &=& \left[ \sinh(\mu_a - \lambda_b) \sinh(\mu_a - \lambda_b + i\zeta)\right]^{-1}, \hspace{1cm}
L^{(n)}_{ab} = \frac{d}{d\lambda} K^{(n)}_{ab}, \nonumber \\
N^{(n)} &=& F^{(n)}_{0} F^{(n)}_1 \frac{\prod_{j=2}^{n} G^{(n)}_{j}}{G^{(n)}_n}.
\label{Hm2P_red}
\end{eqnarray}
in which we have extended the definitions of $\lambda_b$ in (\ref{string_rapidities})
to include $a = 0$ and $a = n+1$ elements for notational convenience.
In the last formula above, the product is defined as unity for $n = 1$.  
Note that these remain $M$-dimensional matrices (the dimension
doesn't change here, in contrast to the Gaudin case).  A similar procedure can be employed to treat the other form factor:
the first $M-1$ columns of the reduced transverse form factor
matrix $H^{-(r)}$ are given by the same expressions as for $H^{(r)}$(bearing in mind that the set $\{ \lambda \}$ then has $M-1$ elements), and
the $M$-th column remains unchanged from equation (\ref{Hm}) in the presence of strings in the set $\{ \lambda \}$.

Therefore, if $\{ \lambda \}$ contains strings and $\{ \mu \}$ does not, we find that the string deviations $\delta$ cancel out
in the product for the longitudinal form factor, and that (with exponentially small corrections of order $\delta$) we can write
\begin{eqnarray}
|\langle \{\mu\}|S_q^z|\{\lambda\}\rangle|^2 
&=& \frac{N}{4} \delta_{q, q_{\{\lambda\}} - q_{\{\mu\}}}
\prod_{j=1}^M \left|\frac{\sinh(\mu_j - i\zeta/2)}{\sinh(\lambda_j - i\zeta/2)}\right|^2
\prod_{j \neq k = 1}^M \left|\sinh(\mu_j - \mu_k + i\zeta)\right|^{-1} \times \hspace{4cm}\nonumber \\
&&\times \prod_{j \neq k = 1, \lambda_j \neq \lambda_k - i\zeta}^M \left|\sinh(\lambda_j - \lambda_k + i\zeta)\right|^{-1}
\frac{\left|\det[{\bf H}^{(r)}(\{\mu\}, \{\lambda\}) - 2{\bf P}^{(r)}(\{\mu\}, \{\lambda\})]\right|^2}
{\left|\det {\bf \Phi} (\{\mu\}) \det {\bf \Phi}^{(r)} (\{\lambda\}) \right|}
\end{eqnarray}
The transverse form factor similarly reduces to
\begin{eqnarray}
|\langle \{\mu\}|S_q^-|\{\lambda\}\rangle|^2 
&=& N \delta_{q, q_{\{\lambda\}} - q_{\{\mu\}}}
|\sin \zeta | \frac{\prod_{j=1}^{M} \left|\sinh(\mu_j - i\zeta/2) \right|^2}{\prod_{j=1}^{M-1}\left|\sinh(\lambda_j - i\zeta/2)\right|^2}
\prod_{j \neq k = 1}^{M} \left|\sinh(\mu_j - \mu_k + i\zeta)\right|^{-1} \times \nonumber \\
&&\times \prod_{j \neq k = 1, \lambda_j \neq \lambda_k - i\zeta}^{M-1} \left|\sinh(\lambda_j - \lambda_k + i\zeta)\right|^{-1}
\frac{\left|\det {\bf H}^{-(r)}(\{\mu\}, \{\lambda\})\right|^2}
{\left|\det {\bf \Phi} (\{\mu\}) \det {\bf \Phi}^{(r)} (\{\lambda\}) \right|}
\end{eqnarray}
For proper string states, all the above functions and determinants are nondegenerate.  
In the next section, we use these results to numerically compute the form factors, and sum them to obtain the dynamical
correlation functions.

\section{Dynamical spin-spin correlations}
\label{Plots}
The strategy to follow is now clear.  We compute the $S^{zz}$ and $S^{-+}$ structure factors by directly summing
the terms on the right-hand side of equation (\ref{S_zz_ff}) 
over a judiciously chosen subset of eigenstates.  The momentum delta functions are broadened to width $\epsilon
\sim 1/N$ using $\delta_{\epsilon} (x) = \frac{1}{\sqrt{\pi}\epsilon} e^{-x^2/\epsilon^2}$ in order to obtain smooth
curves.  

We scan through the eigenstates in the following order.  First, we observe that the form factors of the spin
operators between the ground state and an eigenstate $\{ \lambda \}$ are extremely rapidly decreasing functions of 
the number of holes that need to be inserted in the configuration of the lowest-energy state (in the same base)
in order to obtain the
configuration $\{ I \}$ corresponding to $\{ \lambda \}$.  We therefore scan through all bases and configurations 
for increasing number of holes, starting from one-hole states for $S^{zz}$, and zero-hole states for $S^{-+}$.
Although the number of possible configurations for fixed base and number of holes is a rapidly increasing function
of the number of holes, we find that the total contributions for fixed bases also rapidly decrease for increasing hole numbers.
We therefore limit ourselves to states with up to three holes, corresponding to up to six-particle excitations.

\begin{figure*}
\begin{tabular}{ll}
\includegraphics[width=7.5cm]{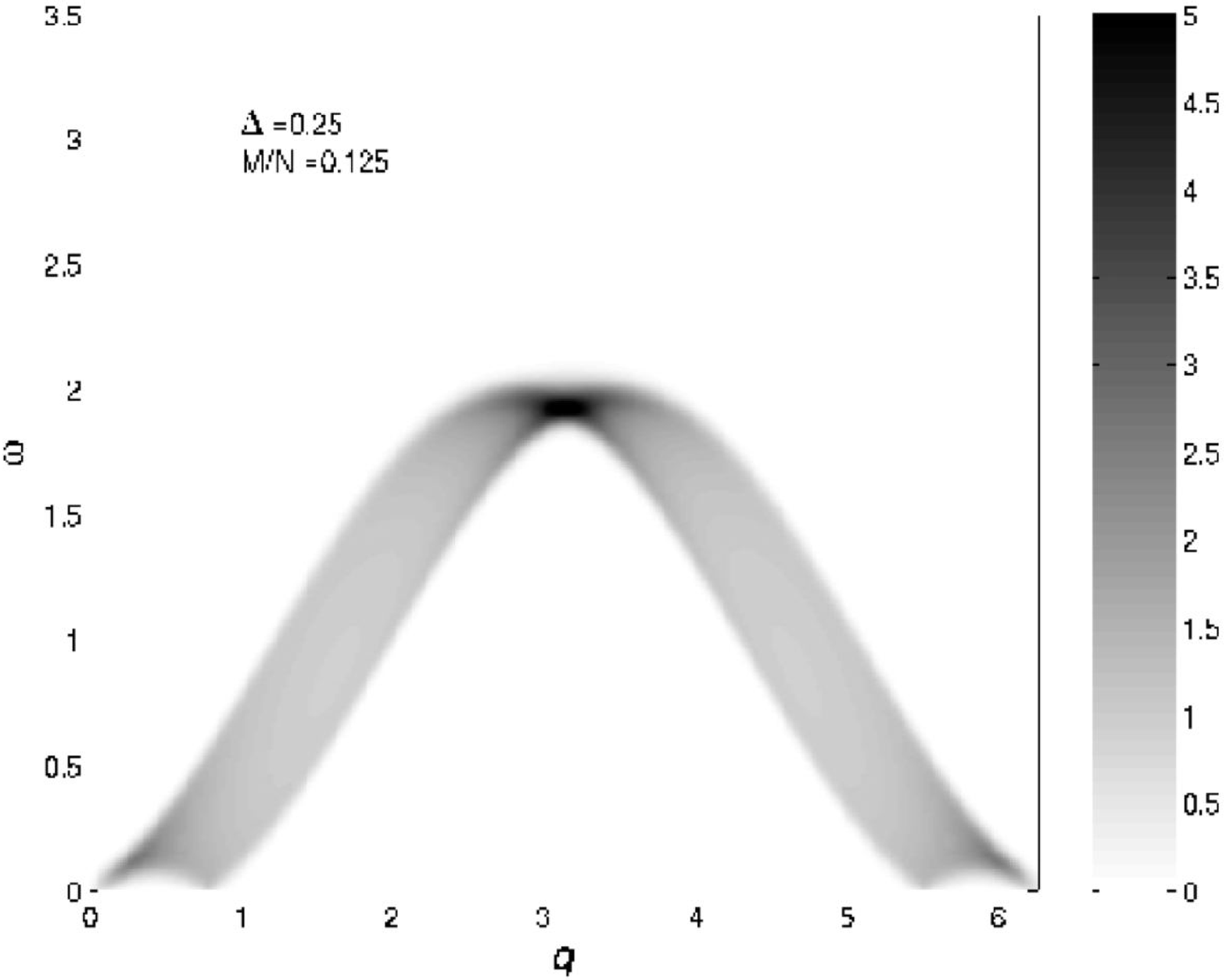} & 
\includegraphics[width=7.5cm]{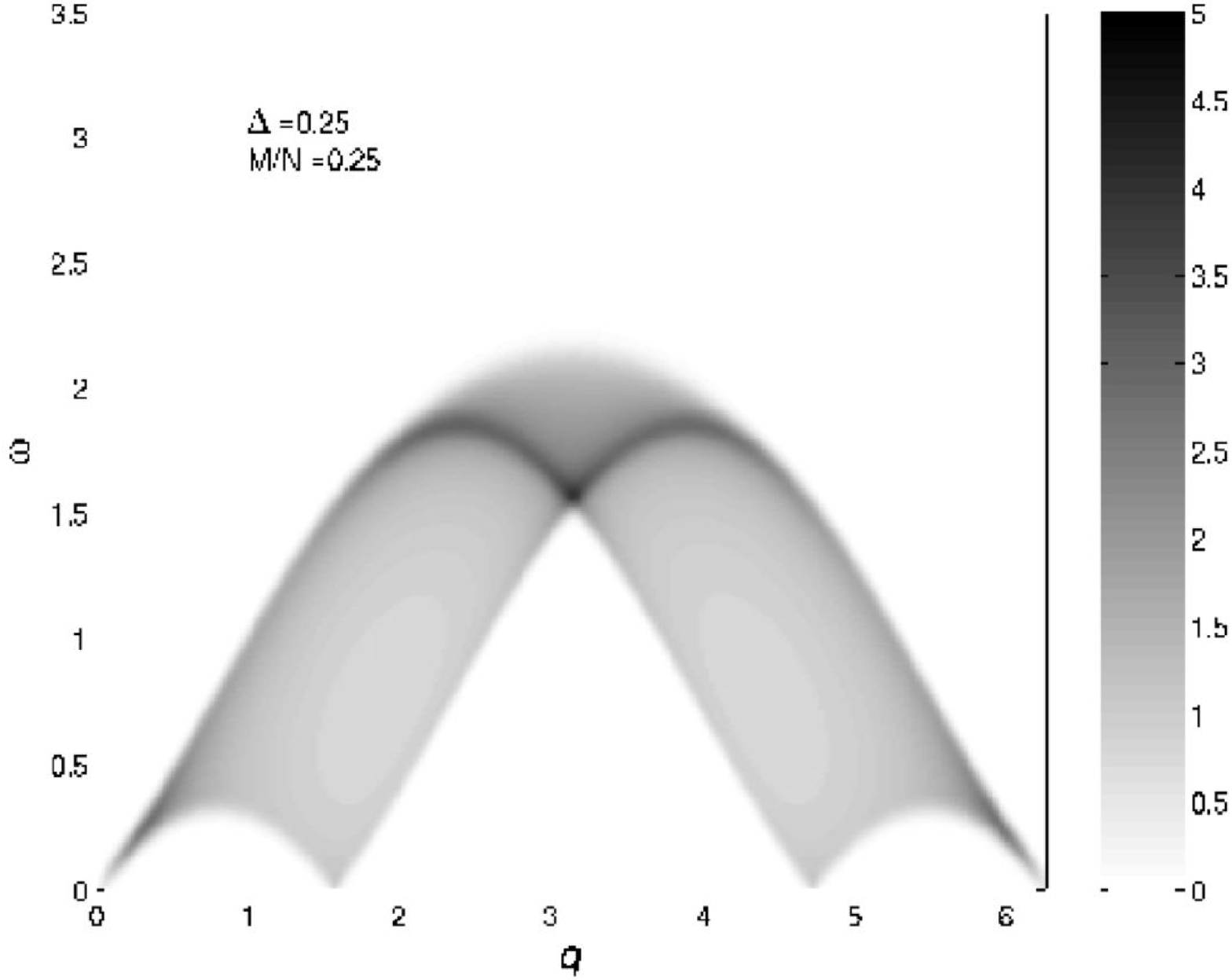} \\ 
\includegraphics[width=7.5cm]{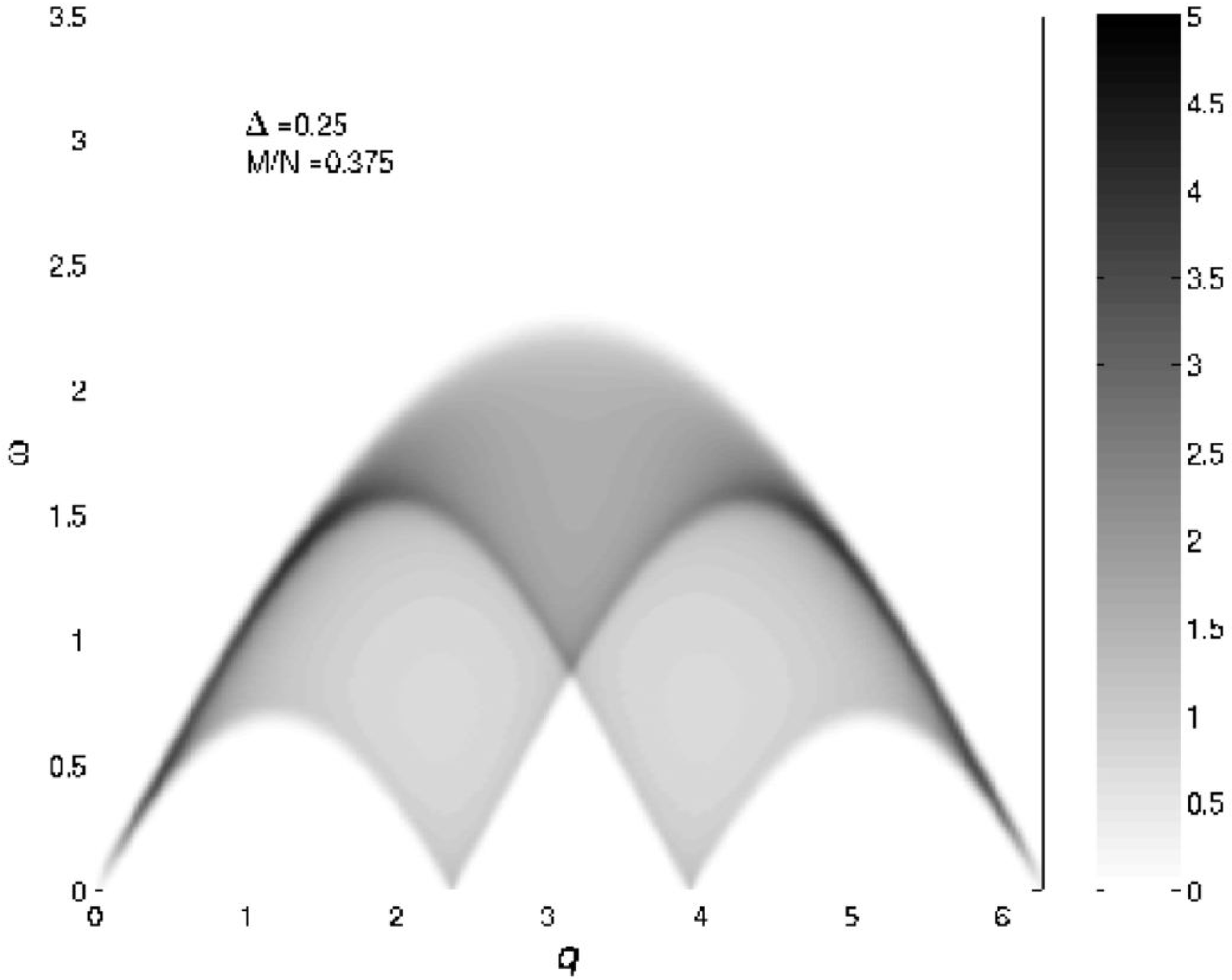} & 
\includegraphics[width=7.5cm]{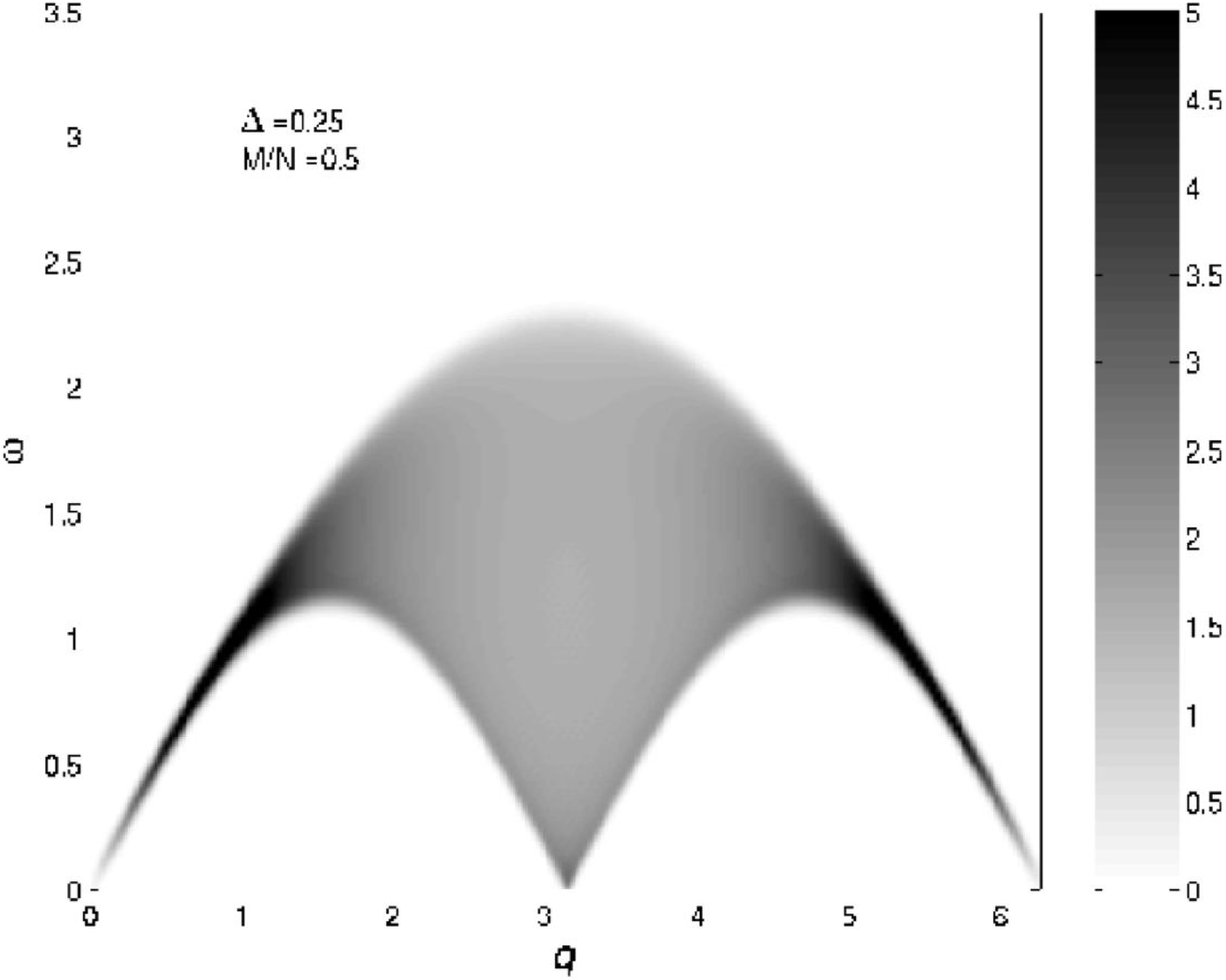}
\end{tabular}
\caption{Longitudinal structure factor as a function of momentum $q$ and frequency $\omega$, for $\Delta = 0.25$, and
$M = N/8, N/4, 3N/8$,  and $N/2$.  Here, $N = 200$ and all contributions up to two holes are taken into account.  
The sum rule is thereby saturated to 98.6\%, 97.2\%, 97.0\% and 99.8\%.}
\label{SZZ_D_0.25}
\end{figure*}

\begin{figure*}
\begin{tabular}{ll}
\includegraphics[width=7.5cm]{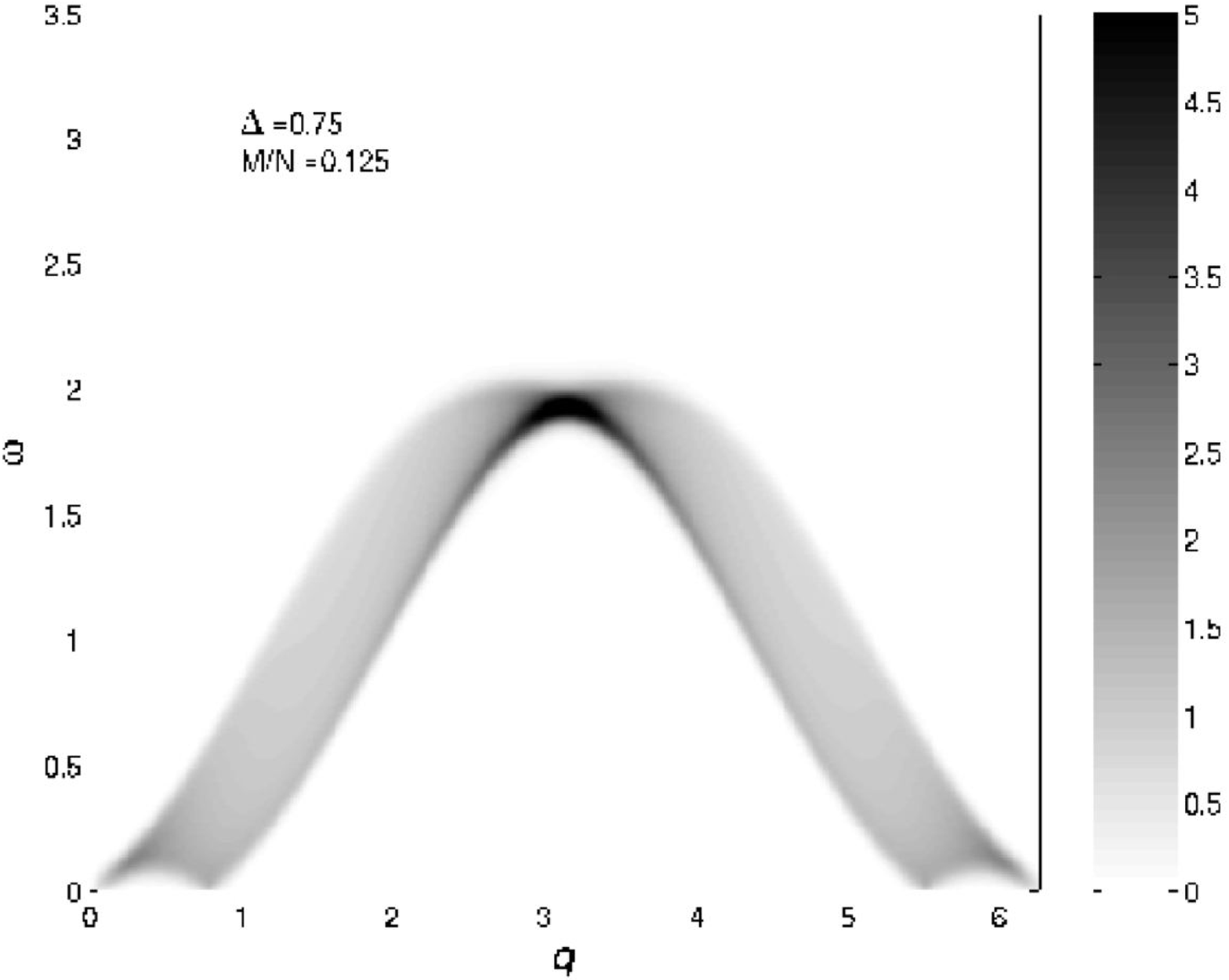} & 
\includegraphics[width=7.5cm]{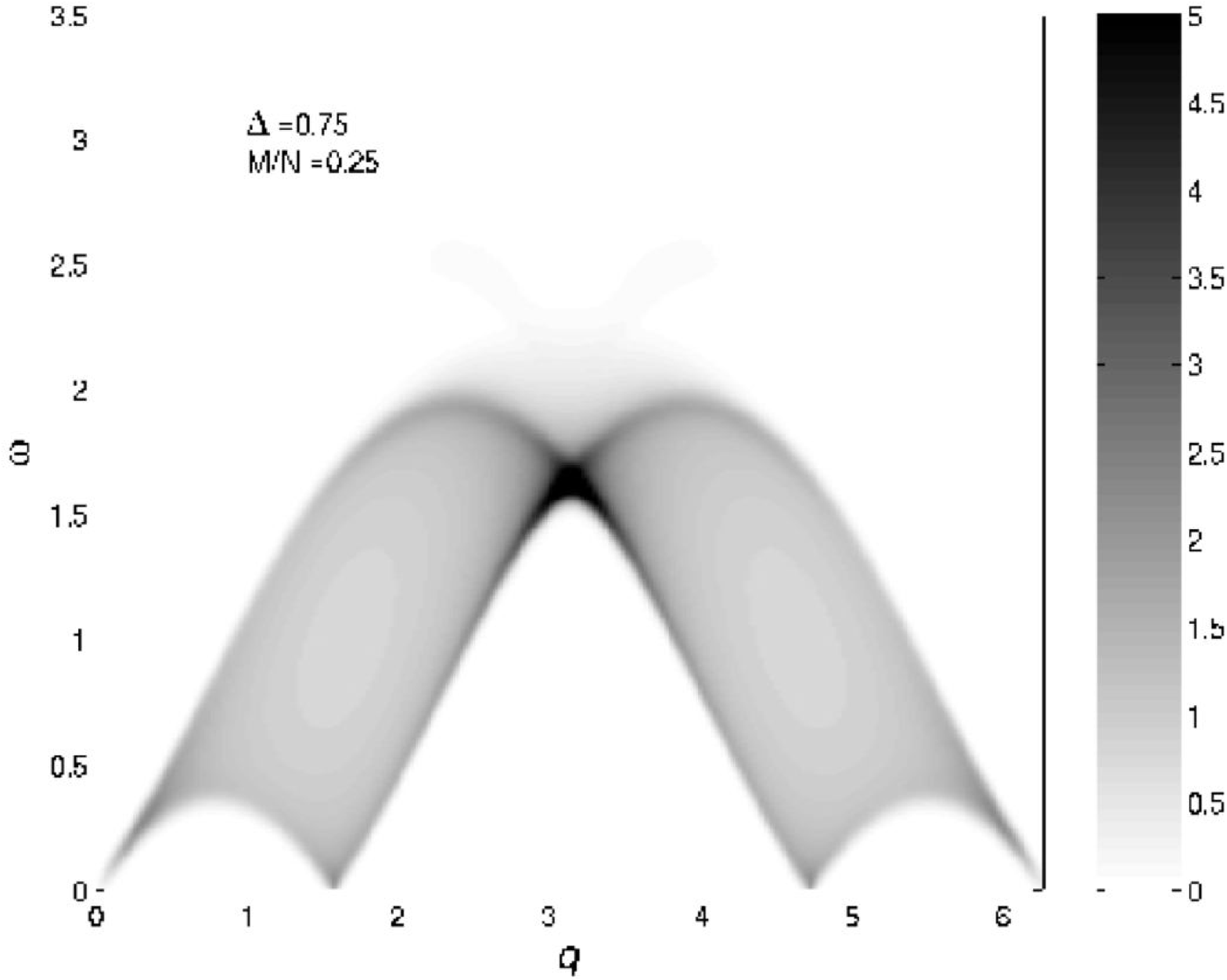} \\ 
\includegraphics[width=7.5cm]{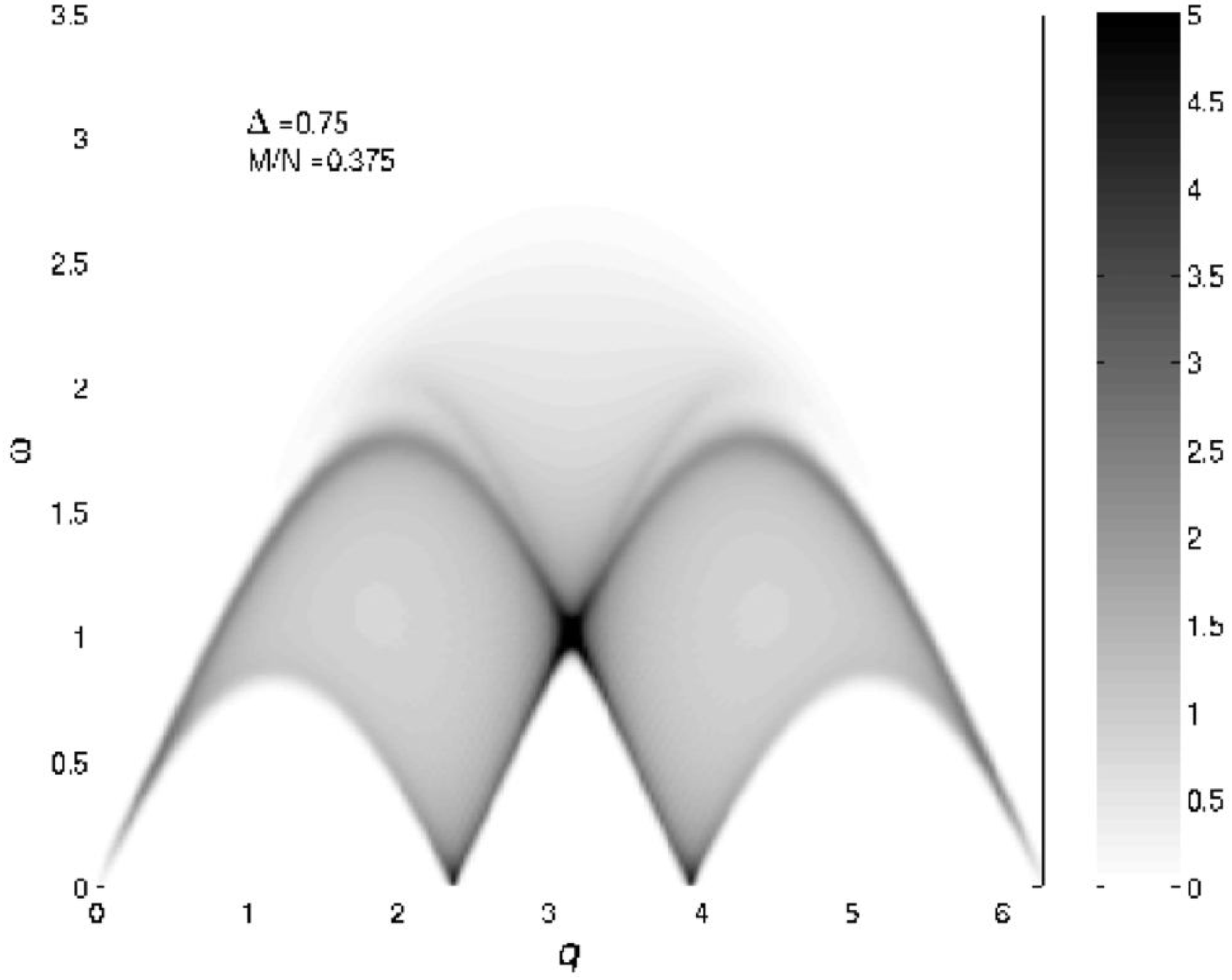} & 
\includegraphics[width=7.5cm]{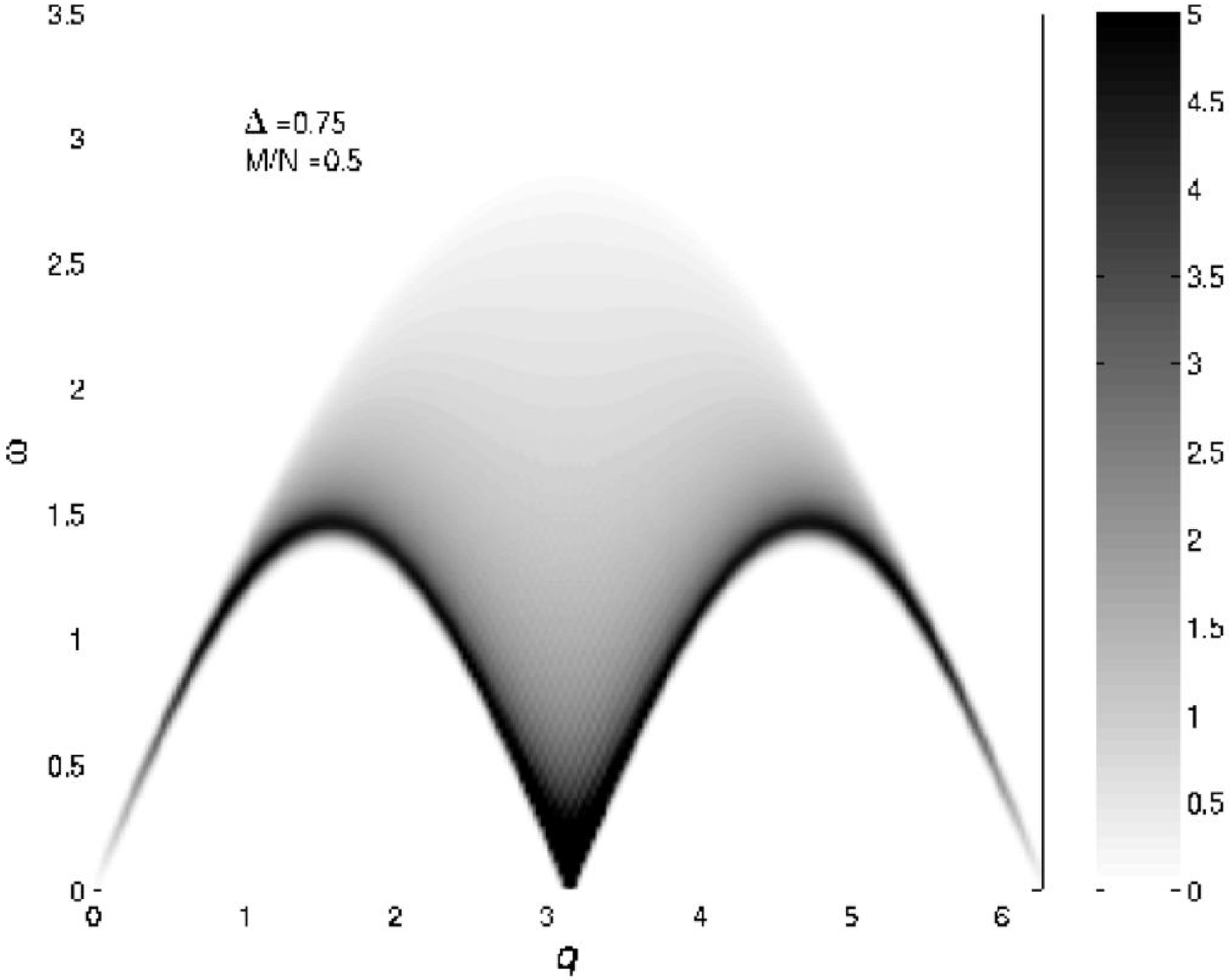}
\end{tabular}
\caption{Longitudinal structure factor as a function of momentum $q$ and frequency $\omega$, for $\Delta = 0.75$, and
$M = N/8, N/4, 3N/8$,  and $N/2$.  Here, $N = 200$ and all contributions up to two holes are taken into account.  
The sum rule is thereby saturated to 98.6\%, 97.0\%, 95.4\% and 97.8\%. }
\label{SZZ_D_0.75}
\end{figure*}

\begin{figure*}
\begin{tabular}{ll}
\includegraphics[width=7.5cm]{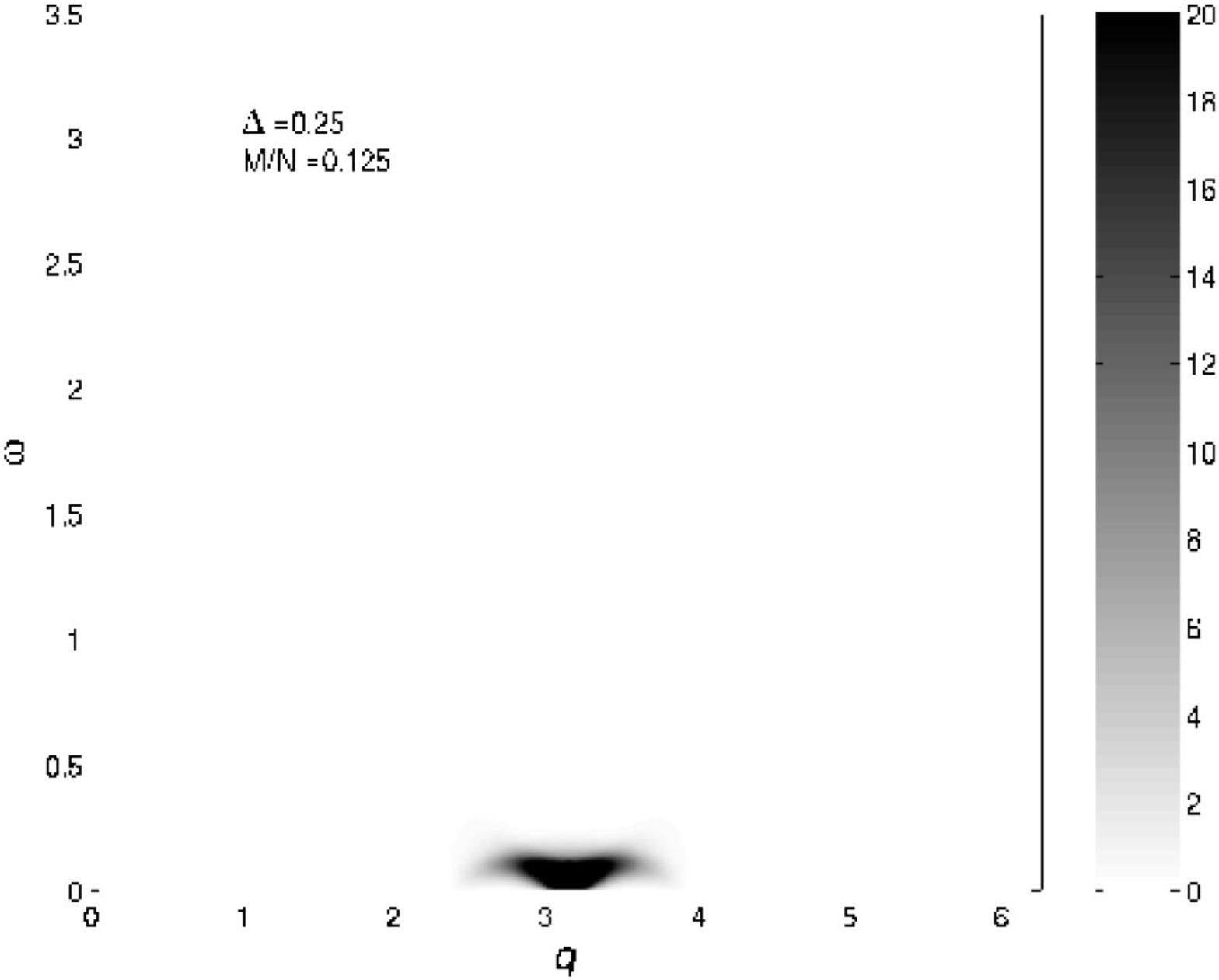} & 
\includegraphics[width=7.5cm]{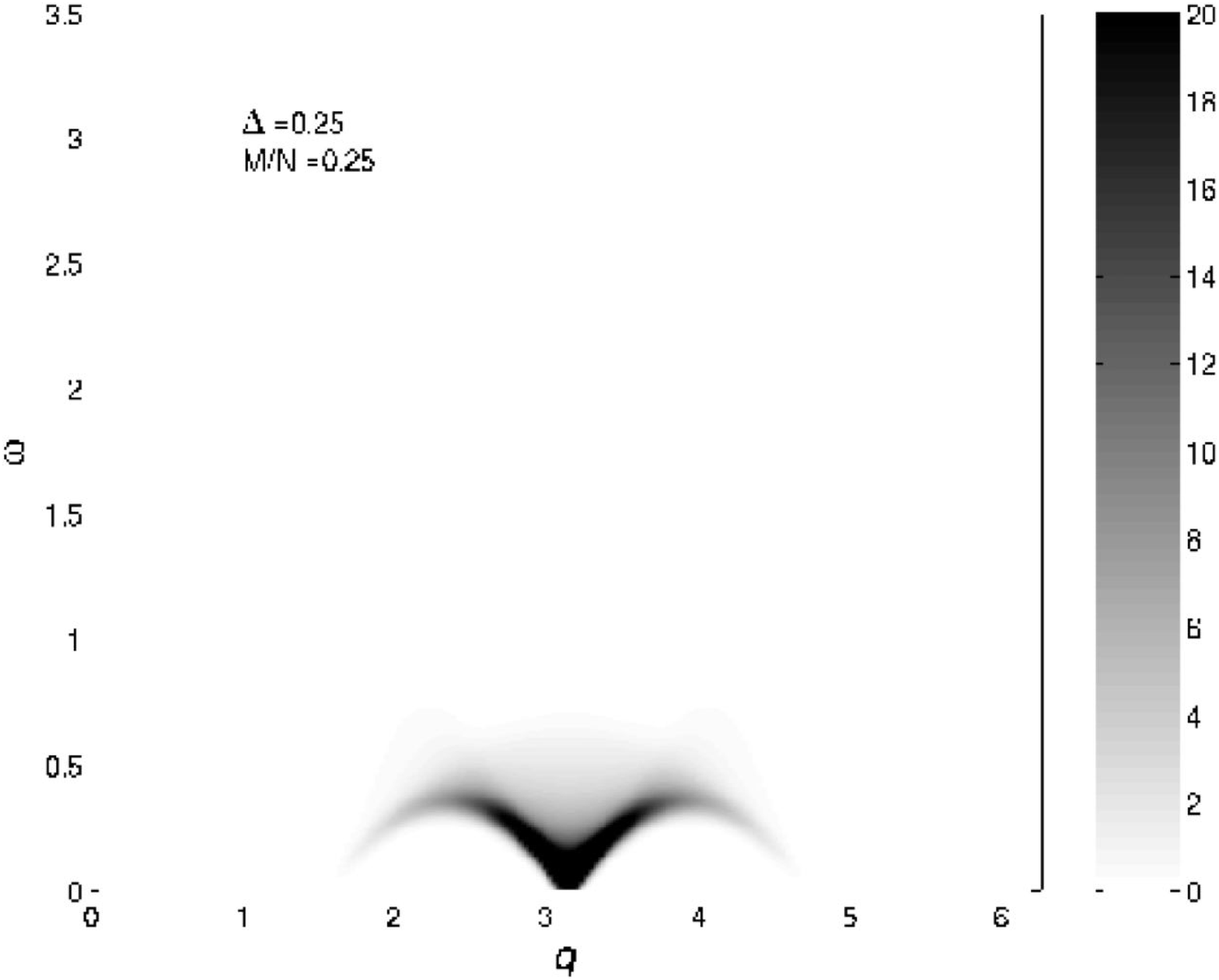} \\ 
\includegraphics[width=7.5cm]{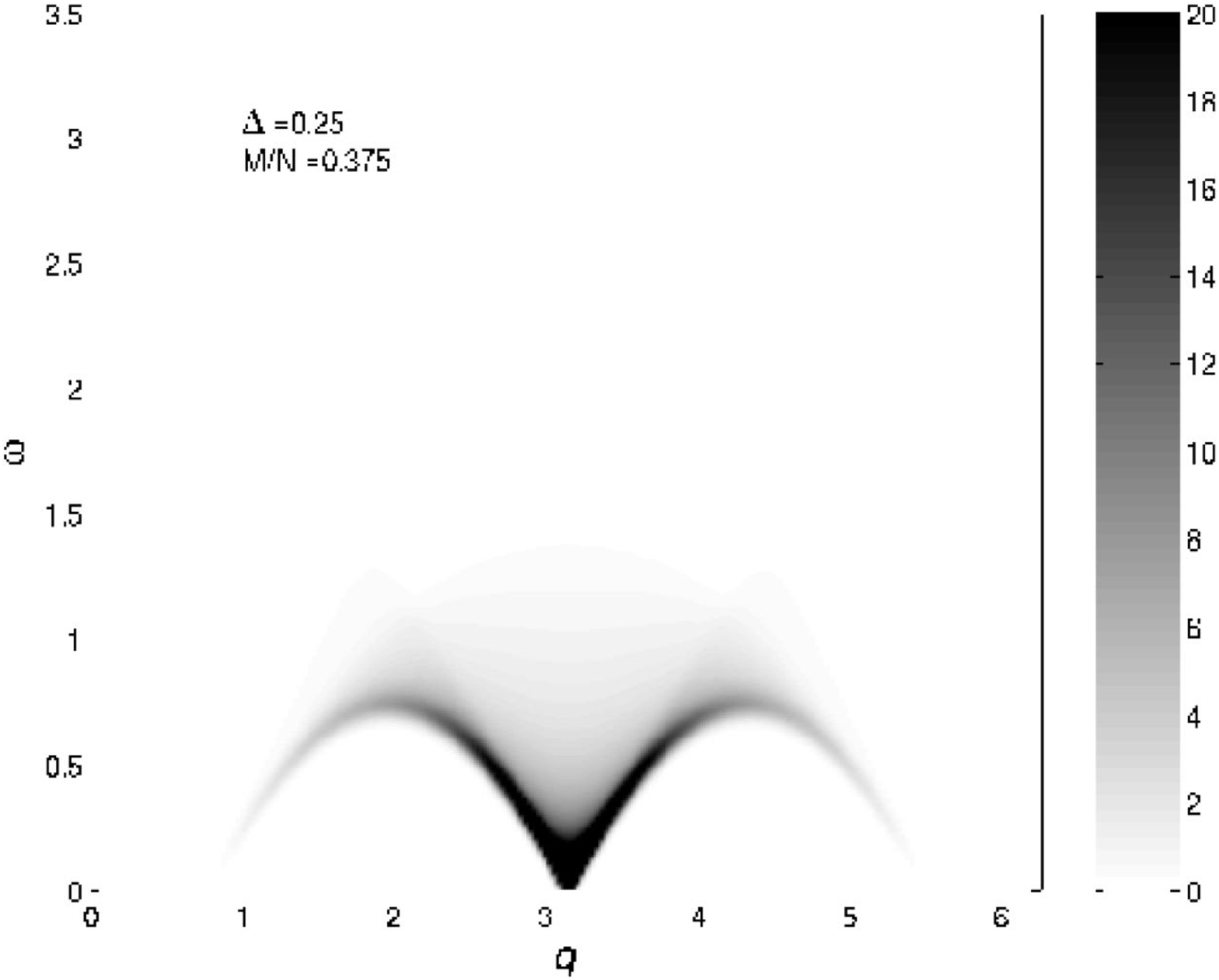} & 
\includegraphics[width=7.5cm]{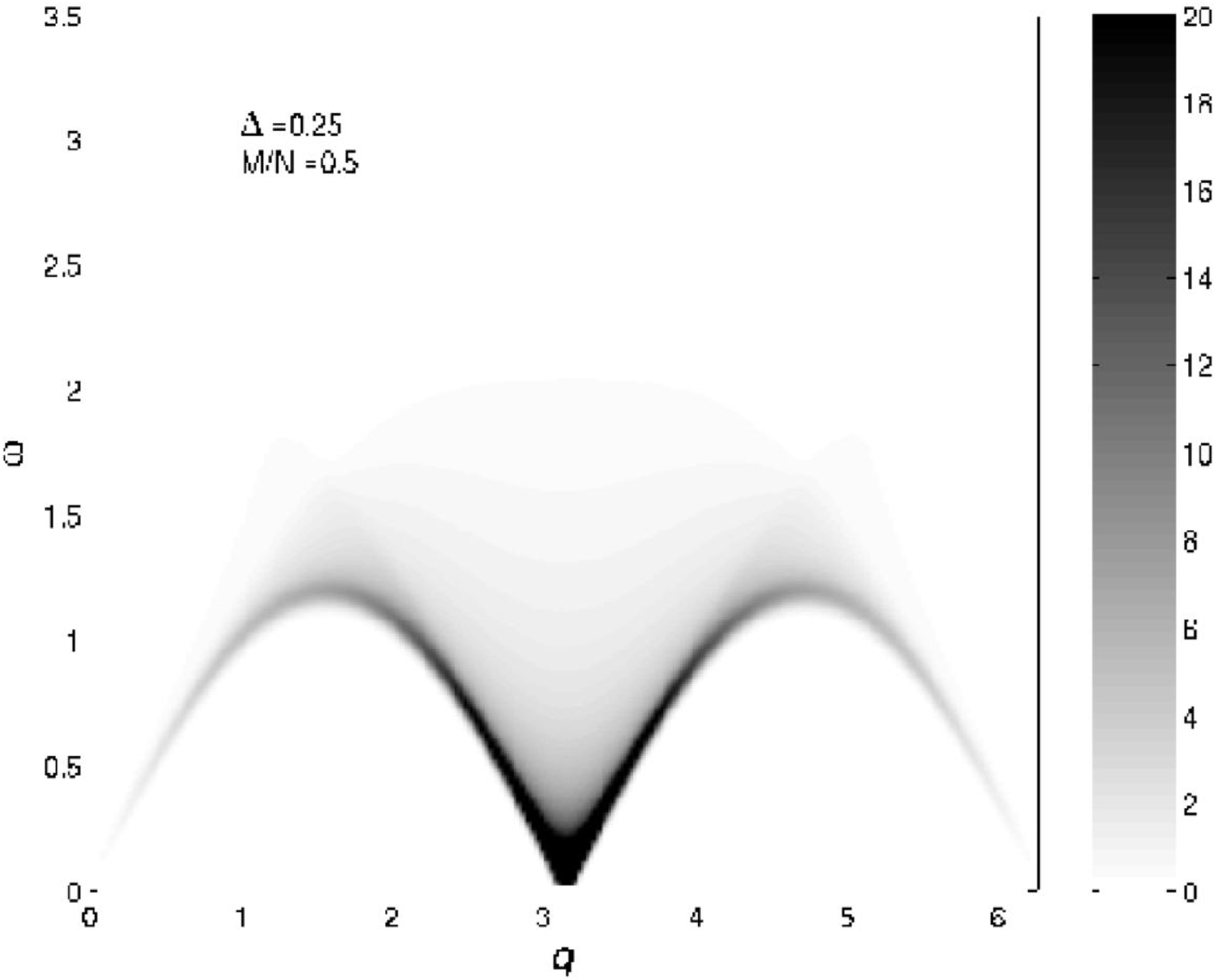}
\end{tabular}
\caption{Transverse structure factor as a function of momentum $q$ and frequency $\omega$, for $\Delta = 0.25$, and
$M = N/8, N/4, 3N/8$,  and $N/2$.  Here, $N = 200$ and all contributions up to two holes are taken into account.  
The sum rule is thereby saturated to 99.3\%, 97.9\%, 95.9\% and 94.6\%. }
\label{SMP_D_0.25}
\end{figure*}

\begin{figure*}
\begin{tabular}{ll}
\includegraphics[width=7.5cm]{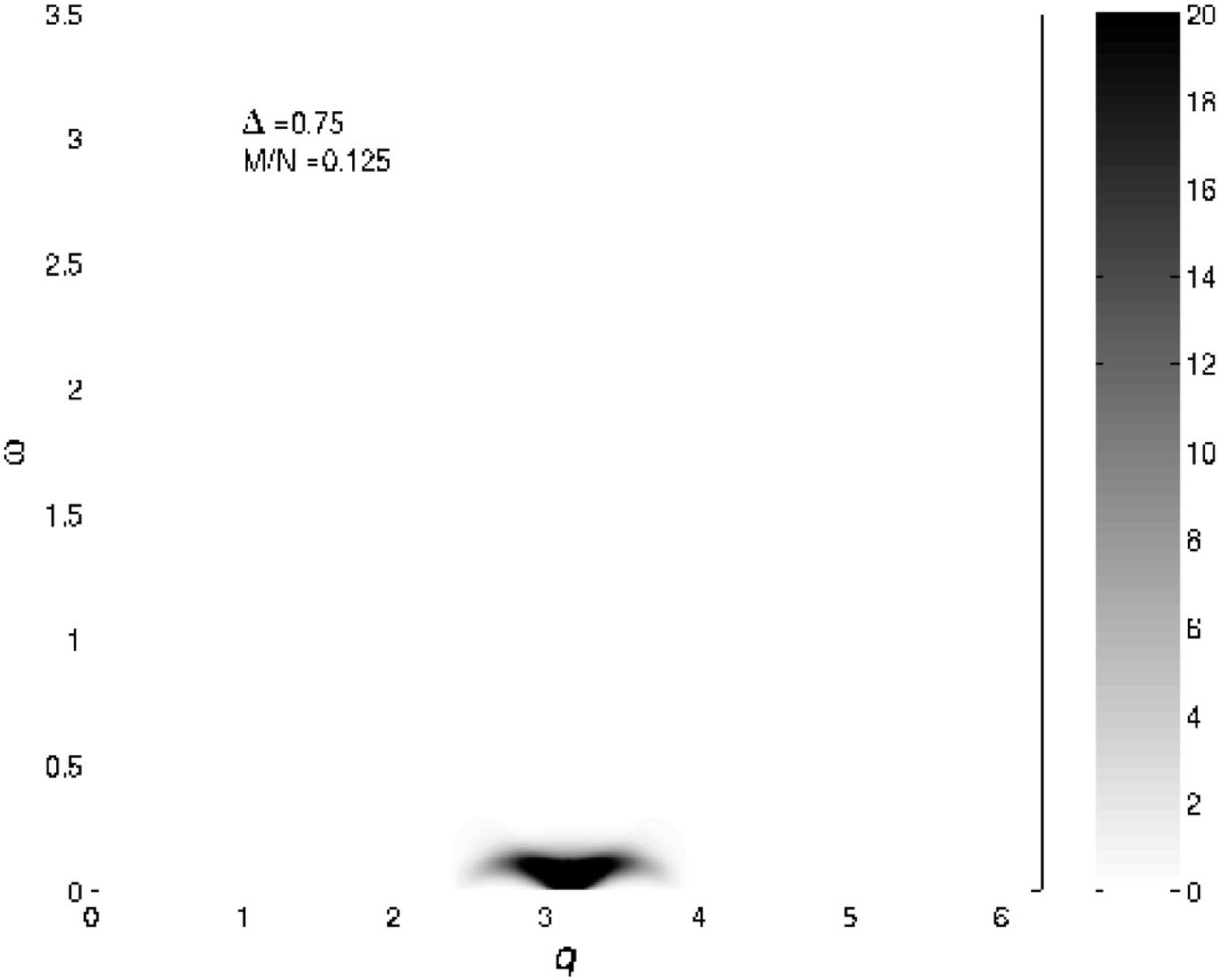} & 
\includegraphics[width=7.5cm]{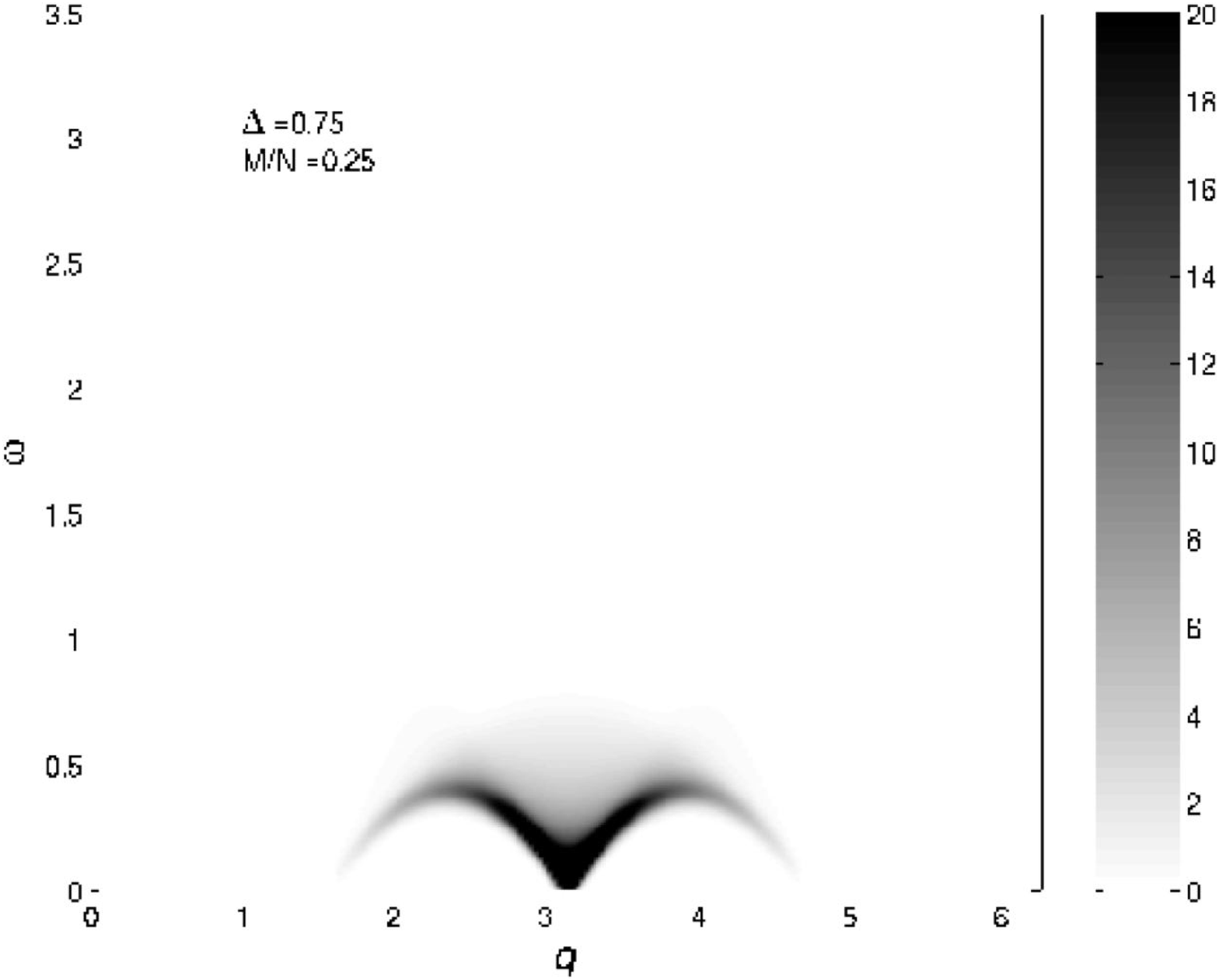} \\ 
\includegraphics[width=7.5cm]{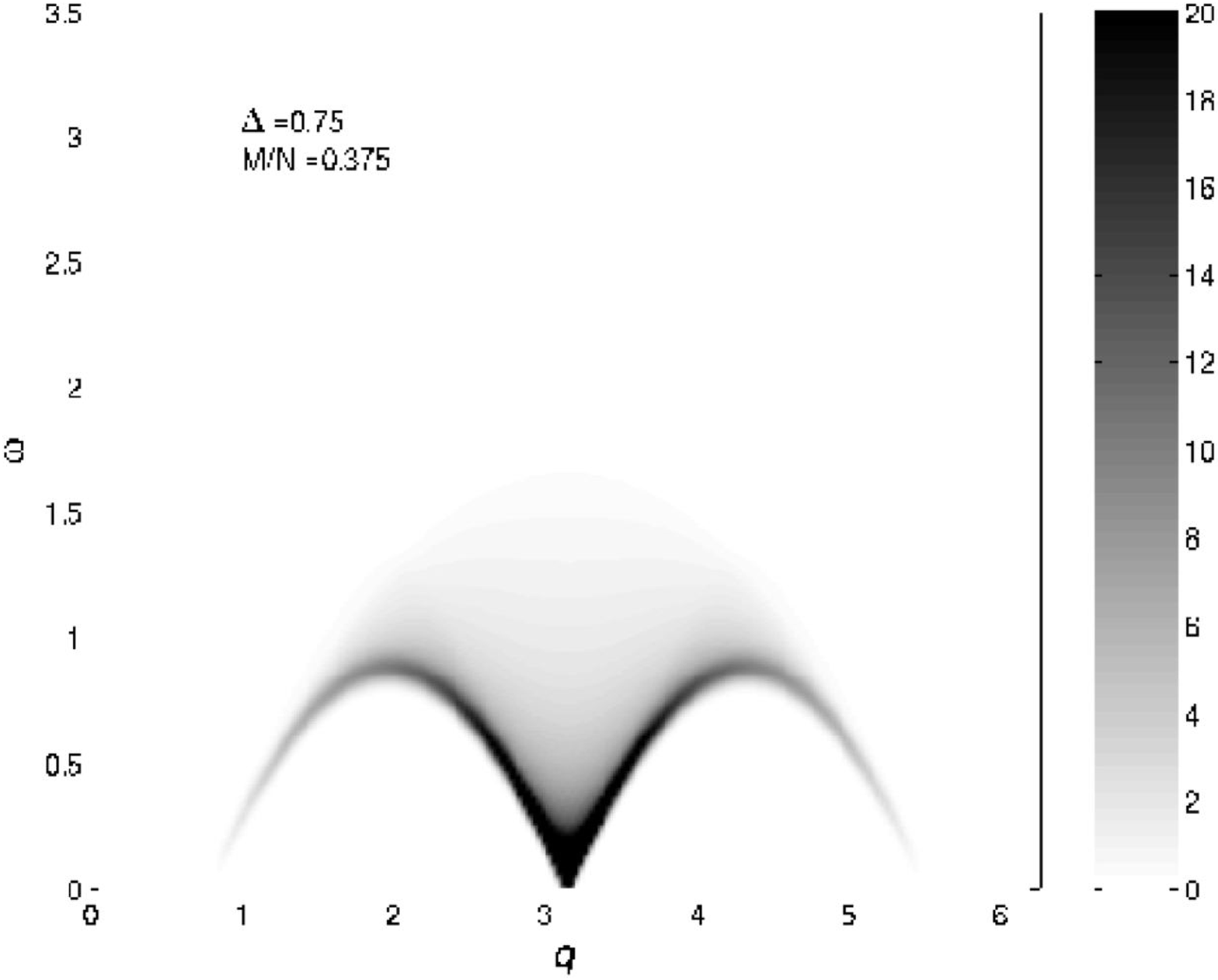} & 
\includegraphics[width=7.5cm]{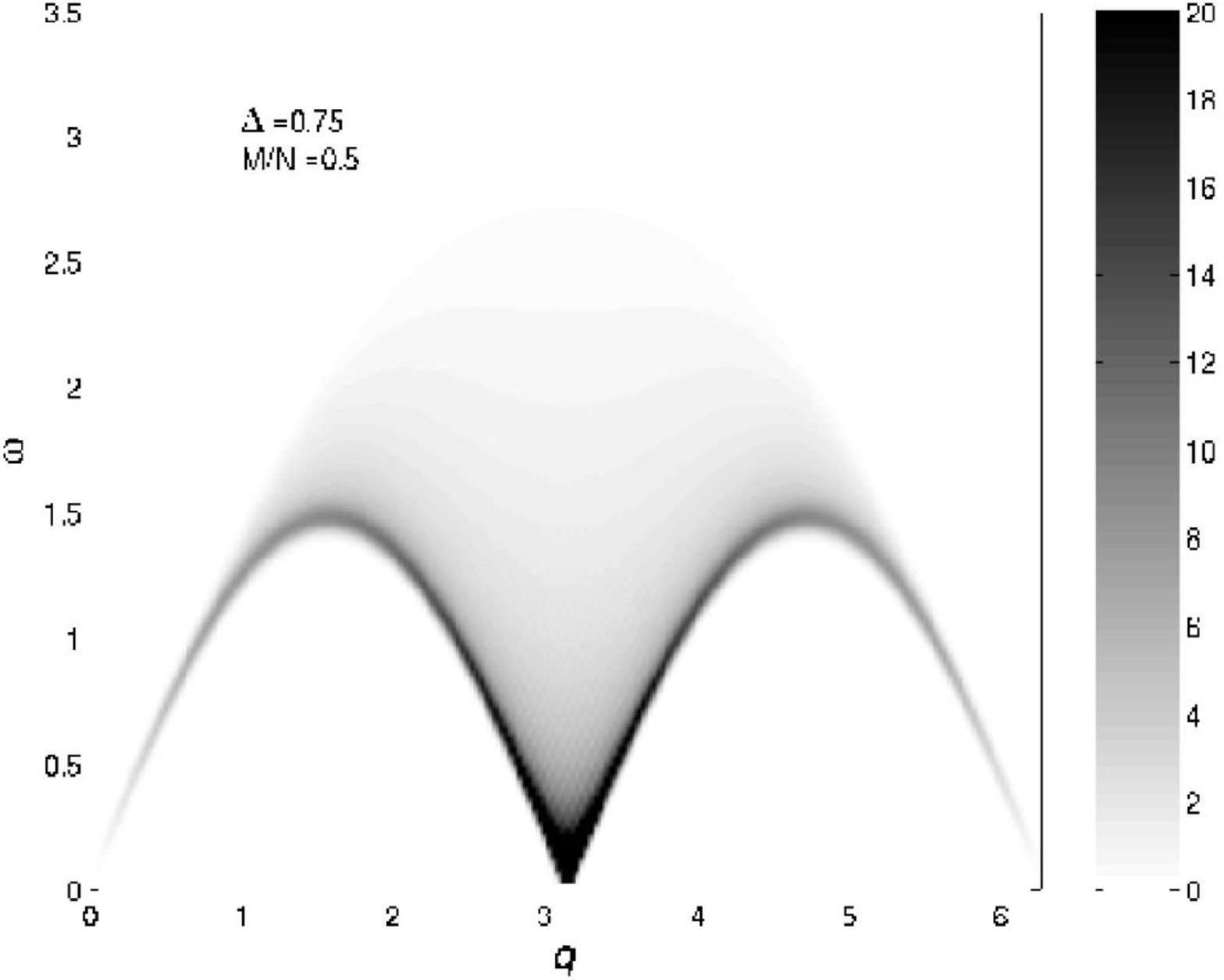}
\end{tabular}
\caption{Transverse structure factor as a function of momentum $q$ and frequency $\omega$, for $\Delta = 0.75$, and
$M = N/8, N/4, 3N/8$,  and $N/2$.  Here, $N = 200$ and all contributions up to two holes are taken into account.  
The sum rule is thereby saturated to 99.3\%, 97.8\%, 96.5\% and 98.8\%. }
\label{SMP_D_0.75}
\end{figure*}

\begin{figure*}
\begin{tabular}{ll}
\includegraphics[width=7.5cm]{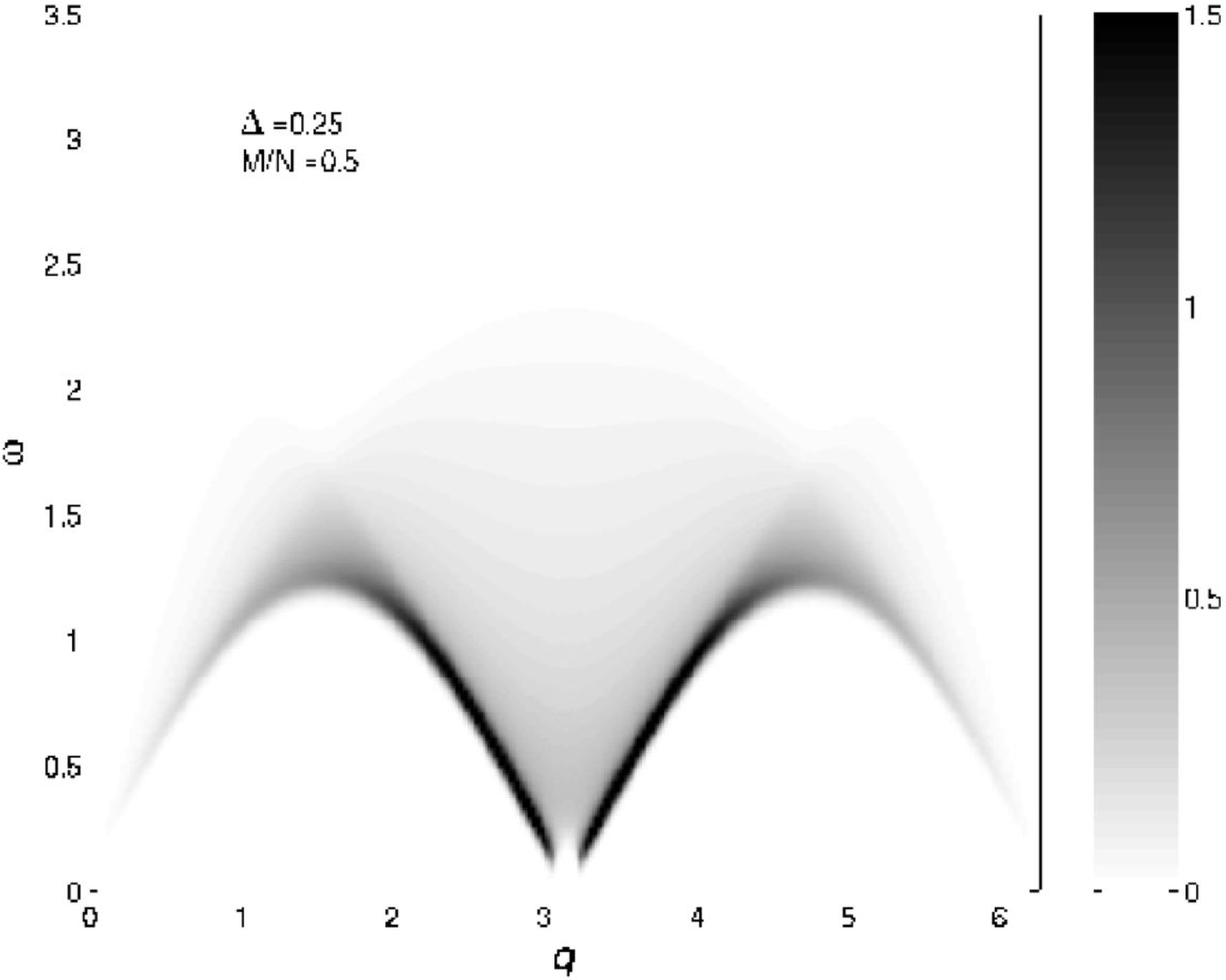} &
\includegraphics[width=7.5cm]{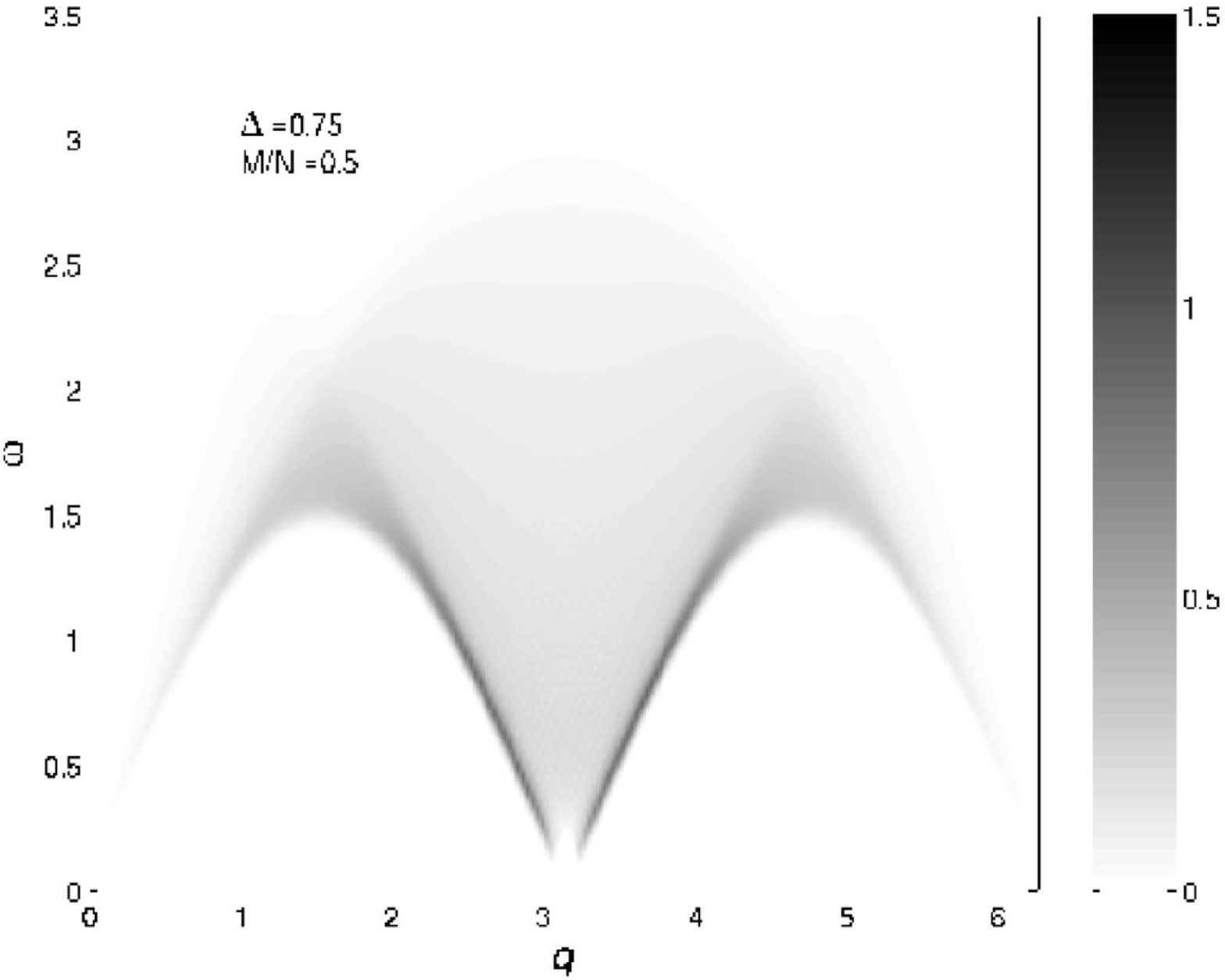}
\end{tabular}
\caption{The two-string contributions to the transverse structure factor at zero magnetic field, 
as a function of momentum $q$ and frequency $\omega$, and for anisotropies $\Delta = 0.25$ and $0.75$.  The density scale 
has been enhanced as compared to that used in the previous figures. Here, $N = 200$ and contributions up to three holes are taken into account.  
The sum rule contributions from these states is 7.2 \% and 6.3 \% respectively. }
\label{2string_S_mp}
\end{figure*}

We can quantify the quality of the present computational method by evaluating the sum rules for the 
longitudinal and transverse form factors.  Namely, by integrating over momentum and frequency, we should saturate the values
\begin{eqnarray}
\int_{-\infty}^{\infty} \frac{d\omega}{2\pi} \frac{1}{N}\sum_{q} S^{zz}(q, \omega) = \frac{1}{4} - \langle S^z\rangle^2 
= \frac{1}{4}\left[1 - (1-\frac{2M}{N})^2\right]
\end{eqnarray}
\begin{eqnarray}
\int_{-\infty}^{\infty} \frac{d\omega}{2\pi} \frac{1}{N}\sum_{q} S^{-+}(q, \omega) = \frac{1}{2} - \langle S^z \rangle = \frac{M}{N}.
\end{eqnarray}
In Fig. \ref{SZZ_D_0.25}, we plot the longitudinal structure factor as a function of momentum and frequency
for anisotropy $\Delta = 0.25$, for four values of the magnetization.  Fig. \ref{SMP_D_0.25} contains the transverse
structure factor for the same anisotropy and magnetizations, whereas Figs. \ref{SZZ_D_0.75} and \ref{SMP_D_0.75}
give plots for another value of anisotropy, $\Delta = 0.75$.  

For all intermediate states involving strings, we explicitly check that the deviations from the string hypothesis
are small.  We find in general that states involving strings of length higher that two are admissible solutions to the Bethe 
equations for high enough magnetizations.  At zero field, only two-string states have exponentially small deviations $\delta$,
and all higher-string states must be discarded, consistent with [\onlinecite{WoynarovichJPA15,BabelonNPB220}].

The relative contributions to the structure factors from different bases is very much dependent on the system size,
the anisotropy, and the magnetization.  In general, we find that two- and four-particle contributions are sufficient to saturate
well over 90\% of the sum rules in all cases, for system sizes up to $N = 200$.  Interestingly, however, we find that string
states also contribute noticeably in many cases.  For example, in Fig. \ref{2string_S_mp}, we plot the zero-field transverse
structure factor contributions coming from intermediate states with one string of length two and up to three holes.  
Around six or seven percent of the weight is accounted for by these states, and similar or somewhat lower figures are found
in other cases.  Strings of length higher than two do not contribute significantly.  For example, we find only 
around 5.7e-8 \% of the sum rule from states with one string of length three, for the longitudinal structure factor
for $\Delta = 0.25$ at $M = N/4$ with $N = 128$.  For $\Delta = 0.75$, we find 6.3e-7 \%.  For the transverse correlators,
the figures are 2.3e-12 \% and 3.1e-12 \%.  Even though these numbers would increase if we could go to larger system
sizes, we do not expect them to ever become numerically significant.  

The imperfect saturation of the sum rules that we obtain in general can be ascribed either to higher states in the 
hierarchy which are not included in our partial summations, or states that are in principle included, but which are
rejected in view of their deviations from the string hypothesis.  As the proportion of excluded string states to
allowable ones can be
rather large (ranging anywhere from zero to fifty percent), we believe the latter explanation to be the correct one.
We will attempt to include these non-string states in a future improvement of our method.

\section{Real space- and time-dependent correlators}
\label{Fourier}
From our results covering the whole Brillouin zone and frequency space, it is straightforward to obtain space-time dependent
correlation functions by inverse Fourier transform:
\begin{eqnarray}
\langle S^a_{j+1} (t) S^{\bar{a}}_1(0) \rangle_c = \frac{1}{N} \sum_{\alpha} |\langle GS | S^a_{q_{\alpha}} | \alpha \rangle |^2
e^{-i q_{\alpha} j - i\omega_{\alpha} t}.
\end{eqnarray}
In the present section, we compare these results to known exact results for equal-time correlators at short distance, and
to the large-distance asymptotic form obtained from conformal field theory.  This comparison can only be made at zero field,
where both sets of results are known exactly.  

\subsection{Small distances and equal time}
It is known that correlation functions of spin operators on the lattice can be represented in general as multidimensional integrals
\cite{JimboPLA168,JimboBOOK,JimboJPA29,KitanineNPB567,KitanineNPB641,KitanineJPA35,Boosh0412191}.  
For example, the spin-spin correlation 
function at distance $j$ is exactly represented by a $j$-fold integral.  These are unfortunately rather difficult to evaluate
either analytically or numerically, and exact answers are only known for small $j$ or in the asymptotic limit $j \rightarrow \infty$.
Recently, large simplifications were obtained for integral representations of the emptiness formation probability 
\cite{BoosJPA34,BoosJPA35,BoosNPB658}, and for third-neighbour correlations \cite{SakaiPRE67}
in the case of the isotropic Heisenberg chain in zero field.  Results for the anisotropic chain in zero field were subsequently obtained in
[\onlinecite{KatoJPA36,TakahashiJPSJ73,KatoJPA37}], yielding exact results for equal-time correlators of spin operators on up to four
adjacent sites.   

Explicitly, these correlators read \cite{KatoJPA37}
\begin{eqnarray}
\langle S^z_j S^z_{j+1} \rangle &=& F^{++}_{++}
 - \frac{1}{4}, \hspace{0.5cm}
\langle S^-_j S^+_{j+1} \rangle = F^{+-}_{-+}
, \hspace{0.5cm}
\langle S^z_j S^z_{j+2} \rangle = 2 F^{+++}_{+++}
-2 F^{++}_{++}
 + \frac{1}{4}, \hspace{0.5cm}
\langle S^-_j S^+_{j+2} \rangle = 2 F^{++-}_{-++}
, \nonumber \\
\langle S^z_j S^z_{j+3} \rangle &=& 2 F^{++++}_{++++}
-2 F^{+-+-}_{-+-+}
 - 3 F^{++}_{++}
+ \frac{3}{4}, \hspace{0.5cm}
\langle S^-_j S^+_{j+2} \rangle = 2 F^{+++-}_{-+++}
+ 2 F^{++--}_{-+-+}
,
\end{eqnarray}
where all blocks $F$, representing correlation functions of $2 \times 2$ fundamental matrices composed of vanishing
elements except for one, are given explicitly as polynomials of the two one-dimensional integrals
\begin{eqnarray}
\zeta_n(j) = \int_{-\infty -i\pi/2}^{\infty - i\pi/2} dx \frac{1}{\sinh x} \frac{\cosh \eta x}{\sinh^j \eta x}, \hspace{1cm}
\zeta^{\prime}_n(j) = \int_{-\infty -i\pi/2}^{\infty - i\pi/2} dx \frac{1}{\sinh x} \frac{\partial}{\partial \eta} 
\frac{\cosh \eta x}{\sinh^j \eta x}.
\end{eqnarray}
The explicit form of these polynomials can
be found for the $XXZ$ case in the second appendix of paper [\onlinecite{KatoJPA37}].  We compare these results to our own
values, obtained from Fourier transform of the form factors, in tables \ref{Comparison_D_0.25} and \ref{Comparison_D_0.75}.  
The deviations between the results are overall even smaller than can be expected on the basis of our slight deviations
from perfect sum rule saturation.  

For the particular case of $\Delta = 0.5$, exact results at zero field were also recently obtained up to distances of
eight sites \cite{Kitanineh0506114}.  We compare these results to our form factor calculations in table \ref{Comparison_D_0.5}.

\begin{table}
\caption{Comparison of equal-time correlation functions 
$\langle S^a_j S^{\bar{a}}_{j+l} \rangle$ at zero field for $\Delta = 0.25$ for small distances $l = 1,2,3$.
Subscript $p$ refers to the exact polynomial representation, whereas $ff$ refers to our results
obtained by summing form factors for all states up to three holes, thereby achieving saturation of the sum
rule to 99.88 \% ($S^{zz}$) and 95.61 \% ($S^{-+}$).} 
\begin{tabular}{|c||c|c||c|c|}
\hline
l & $S^{zz}_p $ & $S^{zz}_{ff}$ & $S^{-+}_p $ & $S^{-+}_{ff} $ \\ \hline
1 & -0.113489 & -0.113337 & -0.316807 & -0.311455 \\ \hline
2 & 0.0129789 & 0.0129605 & 0.180965 & 0.180967 \\ \hline
3 & -0.0163964 & -0.0163965 & -0.152364 & -0.152466 \\ \hline
\end{tabular}
\label{Comparison_D_0.25}
\end{table}

\begin{table}
\caption{Comparison of equal-time correlation functions 
$\langle S^a_j S^{\bar{a}}_{j+l} \rangle$ at zero field for $\Delta = 0.75$ for small distances $l = 1,2,3$.
Subscript $p$ refers to the exact polynomial representation, whereas $ff$ refers to our results
obtained by summing form factors for all states up to three holes, thereby achieving saturation of the sum
rule to 98.73 \% ($S^{zz}$) and 98.89 \% ($S^{-+}$).} 
\begin{tabular}{|c||c|c||c|c|}
\hline
l & $S^{zz}_p $ & $S^{zz}_{ff}$ & $S^{-+}_p $ & $S^{-+}_{ff} $ \\ \hline
1 & -0.136265 & -0.136093 & -0.305461 & -0.304906 \\ \hline
2 & 0.043132 & 0.043427 & 0.140331 & 0.140553 \\ \hline
3 & -0.035605 & -0.035911 & -0.117158 & -0.117201 \\ \hline
\end{tabular}
\label{Comparison_D_0.75}
\end{table}

\begin{table}
\caption{Comparison of equal-time longitudinal correlation function 
$\langle S^z_j S^{z}_{j+l} \rangle$ at zero field for $\Delta = 0.5$ for small distances $l = 1,...,8$.
Subscript $e$ refers to the exact multiple integral representation of reference [\onlinecite{Kitanineh0506114}], 
whereas $ff$ refers to our results
obtained by summing form factors for all states up to three holes, thereby achieving saturation of the sum
rule to 99.2 \%.} 
\begin{tabular}{|c||c|c|}
\hline
l & $S^{zz}_e $ & $S^{zz}_{ff}$ \\ \hline
1 & -0.125 & -0.123817 \\ \hline
2 & 0.027344 & 0.027141 \\ \hline
3 & -0.024475 & -0.024380 \\ \hline
4 & 0.010994 & 0.010955 \\ \hline
5 & -0.011084 & -0.011076 \\ \hline
6 & 0.006248 & 0.006246 \\ \hline
7 & -0.006567 & -0.006580 \\ \hline
8 & 0.004153 & 0.004163 \\ \hline
\end{tabular}
\label{Comparison_D_0.5}
\end{table}

\subsection{The scaling limit}
The continuum limit of the anisotropic $XXZ$ chain in the gapless regime $-1 < \Delta \leq 1$ is 
described by a Gaussian conformal field theory \cite{LutherPRB12,KadanoffAP121}.  Expressions for the lattice
Hamiltonian and for local spin operators as asymptotic expansions of local scaling fields were
obtained to subleading order in [\onlinecite{LukyanovPRB59,LukyanovNPB654}], whose results we use here as comparison.

For large distance, the correlators explicitly read
\begin{eqnarray}
\langle S^-_{j+l} (t) S^{+}_j(0) \rangle_c &\sim& (-1)^l\frac{A/2}{(l_+ l_-)^{\frac{\eta}{2}}} \left\{ 1 - \frac{B}{(l_+ l_-)^{\frac{2}{\eta} - 2}} + ... \right\} 
- \frac{\tilde{A}/2}{(l_+ l_-)^{\frac{\eta}{2} + \frac{1}{2\eta}}} \left\{\frac{1}{2}\left(\frac{l_+}{l_-} + \frac{l_-}{l_+}\right) + \frac{\tilde{B}}{(l_+l_-)^{\frac{1}{\eta} - 1}} 
+ ... \right\}, \nonumber \\
\langle S^z_{j+l} (t) S^{z}_j(0) \rangle_c &\sim& -\frac{1}{4\pi^2 \eta l_+ l_-} \left\{ \frac{1}{2}\left(\frac{l_+}{l_-} + \frac{l_-}{l_+}\right) +
\frac{\tilde{B}_z}{(l_+ l_-)^{\frac{2}{\eta} - 2}} \left( 1 + \frac{2-\eta}{4(1-\eta)} \left(\frac{l_+}{l_-} + \frac{l_-}{l_+}\right) + ... \right)\right\}
+ \nonumber \\ 
&&+\frac{(-1)^l A_z/4}{(l_+ l_-)^{\frac{1}{2\eta}}} \left\{ 1 - \frac{B_z}{(l_+ l_-)^{\frac{1}{\eta} - 1}} + ... \right\}
\end{eqnarray}
with $l_{\pm} = l \pm t/\epsilon$.  
To compare with our results, we use a conformal transformation to finite size, 
so that the chiral distances become $l_{\pm} = \frac{N}{\pi}\sin(\frac{\pi}{N} (l\pm t))$.
Here, the anisotropy appears in the parameter $\eta = 1 - \frac{\zeta}{\pi}$, and the amplitudes 
$A, \tilde{A}, A_z, B, \tilde{B}, B_z$ are known exactly for zero field, {\it e.g.}
\begin{eqnarray}
A = \frac{1}{2(1-\eta)^2} \left[ \frac{\Gamma(\frac{\eta}{2-2\eta})}{2\sqrt{\pi}\Gamma(\frac{1}{2-2\eta})}\right]^{\eta}
\exp \left\{ - \int_0^{\infty} \frac{dt}{t} \left( \frac{\sinh(\eta t)}{\sinh t \cosh((1-\eta)t)} - \eta e^{-2t}\right) \right\}
\end{eqnarray}
(the explicit expressions for the other amplitudes, which we do not reproduce here, 
can be found in [\onlinecite{LukyanovNPB654}]).  

\begin{figure*}
\begin{tabular}{ll}
\setlength{\unitlength}{1mm}
\begin{picture}(50,50)
\put(-33,20){\rotatebox{90}{$(-1)^l \langle S^z_0 S^z_l \rangle$}}
\put(-30,0){\includegraphics[width=7cm]{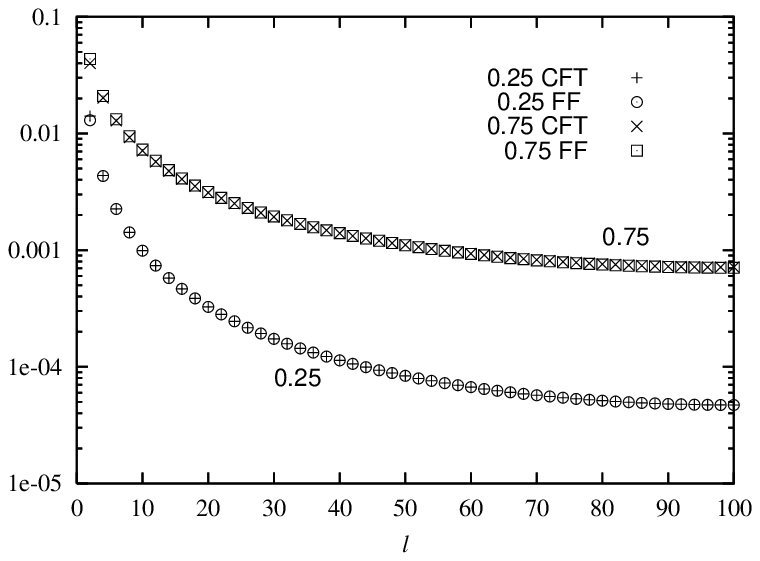}}
\end{picture} & 
\setlength{\unitlength}{1mm}
\begin{picture}(50,50)
\put(0,20){\rotatebox{90}{$(-1)^l \langle S^-_0 S^+_l \rangle$}}
\put(3,0){\includegraphics[width=7cm]{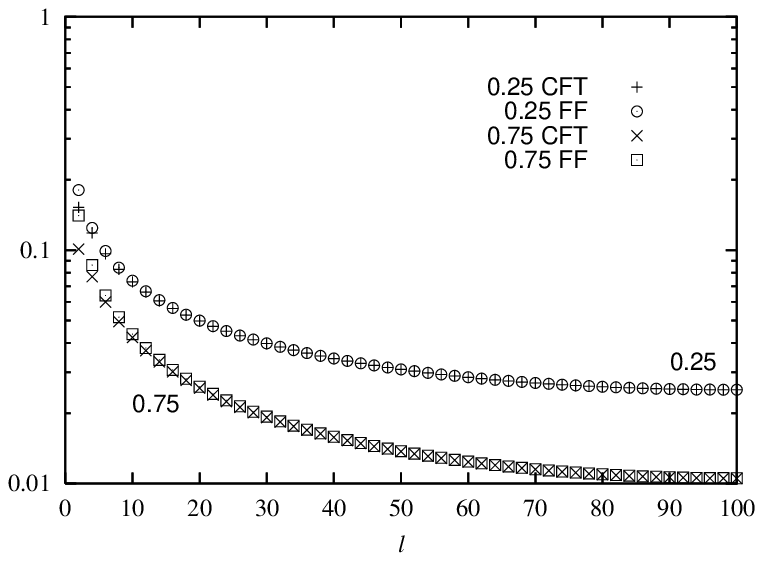}}
\end{picture}
\end{tabular}
\caption{Equal-time spin-spin correlation functions as a function of lattice distance, for anisotropies $\Delta = 0.25$
and $\Delta = 0.75$, 
for chains of $N = 200$ sites.  
Our form factor results (for intermediate states containing up to three holes) are compared to the asymptotics
from CFT adapted to finite size.  The deviations are well within those expected from the sum rules and finite size 
effects for the form factors, and
subleading corrections to the CFT result.  Example numbers are given in tables \ref{Asymptotics_Comparison_D_0.25} and
\ref{Asymptotics_Comparison_D_0.75}.}
\label{Scaling_comparison}
\end{figure*}

Tables \ref{Asymptotics_Comparison_D_0.25} and \ref{Asymptotics_Comparison_D_0.75} contain 
comparisons of the numerical values of the equal-time spin-spin correlation functions 
computed using the CFT aymptotic expansions and our form factor calculations.
Fig. \ref{Scaling_comparison} contains plots of the results as a function of lattice distance.  Once again, the deviations
are even smaller than should be expected, considering the imperfect sum rule saturation the the fact that these remain
finite size comparisons.  The only substantial deviations occur at small distances, where the conformal approach yields
incorrect results (as a comparison to the data from the previous subsection shows).  

\begin{table}
\caption{Comparison of equal-time correlation functions 
$\langle S^a_j S^{\bar{a}}_{j+l} \rangle$ at zero field for $\Delta = 0.25$ for large distances $l$.
Subscript $CFT$ refers to the scaling prediction, whereas $ff$ refers to our results
obtained by summing form factors for $N = 200$, for intermediate states including up to three holes, thereby achieving saturation of the sum
rule to 99.88 \% ($S^{zz}$) and 95.61 \% ($S^{-+}$).} 
\begin{tabular}{|c||c|c||c|c|}
\hline
l & $S^{zz}_{CFT} $ & $S^{zz}_{ff}$ & $S^{-+}_{CFT} $ & $S^{-+}_{ff} $ \\ \hline
10 & 9.89624e-4 & 9.89375e-4 & 7.31376e-2 & 7.39883e-2 \\ \hline
25 & -3.77195e-4 & -3.77096e-4 & -4.41637e-2 & -4.40758e-2 \\ \hline
40 & 1.13604e-4 & 1.13539e-4 & 3.42939e-2 & 3.43211e-2 \\ \hline
55 & -1.11476e-4 & -1.11424e-4 & -2.95934e-2 & -2.95775e-2 \\ \hline
70 & 5.70852e-5 & 5.70336e-5 & 2.69545e-2 & 2.69692e-2 \\ \hline
85 & -7.21563e-5 & -7.21058e-5 & -2.56504e-2 & -2.56405e-2 \\ \hline
100 & 4.71343e-5 & 4.70918e-5 & 2.52109e-2 & 2.52193e-2 \\ \hline
\end{tabular}
\label{Asymptotics_Comparison_D_0.25}
\end{table}

\begin{table}
\caption{Comparison of equal-time correlation functions 
$\langle S^a_j S^{\bar{a}}_{j+l} \rangle$ at zero field for $\Delta = 0.75$ for large distances $l$.
Subscript $CFT$ refers to the scaling prediction, whereas $ff$ refers to our results
obtained by summing form factors for $N = 200$, for intermediate states including up to three holes, thereby achieving saturation of the sum
rule to 98.73 \% ($S^{zz}$) and 98.89 \% ($S^{-+}$).} 
\begin{tabular}{|c||c|c||c|c|}
\hline
l & $S^{zz}_{CFT} $ & $S^{zz}_{ff}$ & $S^{-+}_{CFT} $ & $S^{-+}_{ff} $ \\ \hline
10 & 7.10534e-3 & 7.26298e-3 & 4.22187e-2 & 4.36235e-2 \\ \hline
25 & -2.49782e-3 & -2.51931e-3 & -2.24156e-2 & -2.21879e-2 \\ \hline
40 & 1.39631e-3 & 1.39725e-3 & 1.57658e-2 & 1.58537e-2 \\ \hline
55 & -1.03764e-3 & -1.03205e-3 & -1.31010e-2 & -1.30427e-2 \\ \hline
70 & 8.26181e-4 & 8.18235e-4 & 1.14797e-2 & 1.15143e-2 \\ \hline
85 & -7.56746e-4 & -7.47886e-4 & -1.08240e-2 & -1.07873e-2 \\ \hline
100 & 7.13747e-4 & 7.04615e-4 & 1.05099e-2 & 1.05363e-2 \\ \hline
\end{tabular}
\label{Asymptotics_Comparison_D_0.75}
\end{table}

\section{Conclusion}
\label{Conclusion}
The existence of determinant representations for lattice form factors of local spin operators makes it possible to 
compute dynamical correlation functions to a high degree of accuracy, and we have here attempted to push the
results of [\onlinecite{Caux0502365}] further by including the remaining spin-spin correlators, 
studying their general field dependence, including contributions from intermediate states containing complex solutions
to the Bethe equations in the presence of strings, and computing explicitly real space- and time-dependent correlators.
Although intermediate states with strings of length two do indeed give non-negligible contributions to the sum rules, 
we have found that strings of any higher length have form factors which are much too small to be of significance. 

We have demonstrated the reliability of our results by performing comparisons 
with exact integral representations for equal time, small-distance correlators at zero field, 
and exact formulae for the large-distance asymptotics (again at zero field).  
We do wish to emphasize that the present method, unlike these calculations, is not limited to zero field.  
Also, unlike other numerical methods based on the Density Matrix Renormalization Group (DMRG), 
it is by construction directly applicable to dynamics.  In other words, we are able
to numerically compute all spin-spin correlators either as a function of momentum and frequency, or space and time
separation, for any value of magnetization, for system sizes well beyond the reach of any other numerical method.

\begin{acknowledgments}
J.-S. C. acknowledges support from CNRS and the Stichting voor Fundamenteel Onderzoek der Materie (FOM).
J.-M. M. is supported by CNRS, and acknowledges support from the EUCLID European network.
\end{acknowledgments}

\end{document}